\documentclass[onecolumn,secnumarabic,amssymb, nobibnotes, aps, pre]{revtex4-1}

\usepackage{graphicx}

\begin{document}

\title{Generalized Darboux transformation and higher-order rogue wave solutions  to the Manakov system.}%

\author{Serge P. Mukam$^{1,2,*}$, Victor K. Kuetche$^{1,2,3,4}$ and Thomas B. Bouetou$^{1,2,3,4}$}%
\email[Corresponding author: ]{sergemukam@yahoo.fr}
\affiliation{$^{1}$Ecole Nationale Sup\'{e}rieure Polytechnique,
University of Yaounde I, P.O. Box. 8390, Cameroon}
\affiliation{$^{2}$Department of Physics, Faculty of Science, University
of Yaounde I, P.O. Box. 812, Cameroon}
\affiliation{$^{3}$Centre d'Excellence en Technologies de l'Information et de la Communication (CETIC), University of Yaounde I, P.O. Box 812, Yaounde, Cameroon}
\affiliation{$^{4}$The Abdus Salam International Centre for Theoretical Physics (ICTP), Strada Costiera, Trieste 11-I-34151, Italy}
\date{August 10, 2010 }%
\begin{abstract}
In this paper, we construct a generalized recursive Darboux transformation of a focusing  vector nonlinear Schr$\ddot{o}$dinger equation known as the Manakov system. We apply this generalized recursive Darboux transformation to the Lax-pairs of this system in view of generating the Nth-order vector generalization rogue wave solutions with the same spectral parameter through a direct iteration rule. As a result, we discuss the first, second and  third-order vector generalization rogue wave solutions while illustrating these features with some depictions. We show that higher-order rogue wave solutions depend on the values of their free parameters.
\end{abstract}
\maketitle


\section{Introduction}
Understanding the behavior and the dynamics of natural phenomena stand to be worth fundamental. That is why it has been one of the most challenging aspects of modern science and  technology in studying nonlinear nature of system. The importance of nonlinearity has been well-appreciated for many years, because nonlinearity is a fascinating occurrence of nature in the context of large amplitude waves or high-intensity laser pulses observed in various fields. This fascinating subject has branched out in almost all areas of science, and its applications are percolating through the whole science. In general, nonlinear evolution equations exhibiting a wide range of high complexities in terms of different linear and nonlinear effects model nonlinear phenomena. Nonlinear science has experienced an explosive growth by the invention of several exciting and fascinating new concepts in the past few decades, such as solitons, dispersion-managed solitons, dromions, rogue waves, among others \cite{R1,R2,R3,R4}.

Rogue waves, also called freak waves, giant waves or killer waves have attracted considerable attentions. A rogue wave is a large-amplitude local wave, short-lived wave, meet in an ocean that appears from nowhere and disappears without a trace. Rogue waves appear not only in oceanic conditions \cite{R5}, but  also in optics \cite{R6}, superfluids \cite{R7}, Bose-Einstein condensates \cite{R8} and in the form of capillary waves \cite{R9}. In any of these disciplines, new studies of rogue waves enrich the concept and lead to progress toward a comprehensive understanding of this still mysterious phenomenon.
The first-order rational solution for the nonlinear Schr$\ddot{o}$dinger equation (NLSE) was given by Peregrine \cite{R10}. Akhmediev and co-workers have calculated the simplest rogue wave solutions for the NLSE \cite{R11,R12}. The construction of higher-order analogues is actually a challenging problem. The construction of higher-order rogue wave solutions needs a simple approach which is the generalized Dardoux transformation (DT). This approach was proposed by Guo et al \cite{R13}.

The DT, originating from the work of Darboux \cite{R14} on the Sturm-Liouville equation, is a powerful method for constructing solutions for integrable systems such as the NLSE, the Korteweg-de Vries (KdV) equation, the Kadomtsev-Petviashvili equation, the Davey-Stewartson equation  and  the  Toda  lattice  equation just to name a few. But the original DT is not applicable directly to obtain the rogue wave solutions for the nonlinear wave equations \cite{R13}. Matveev \cite{R15} introduced the so-called generalized DT and the positon solutions were calculated for the celebrated KdV equation. Recently, Guo et al. \cite{R13} re-examined Matveev's generalized DT and proposed a new approach to derive the generalized DT for the KdV and the NLS equations.

In this work, we discuss the Guo et al \cite{R13}'s approach to a focusing vector NLSE (VNLSE), known as Manakov system given as follows
\begin{eqnarray}\label{Eq1}
  iu_{t}+\frac{1}{2}u_{xx}+u(|u|^{2}+|v|^{2}) &=& 0,
  \end{eqnarray}
  \begin{eqnarray}\label{Eq2}
  iv_{t}+\frac{1}{2}v_{xx}+v(|u|^{2}+|v|^{2}) &=& 0,
\end{eqnarray}
 with $x$ and $t$ being two independent variables, $u(x,t)$ and $v(x,t)$ standing for complex envelop of two field components. This system has many physical significant applications such as the propagation in elliptically birefringent optical fibers \cite{R16} and for modeling crossing sea waves \cite{R17}.

  We consider in this paper the VNLS equations in the anomalous dispersion regime. The aim of this work is to construct Nth-order rogue wave solutions of the previous system. The main tool is the generalized DT. Based on the Darboux matrix method \cite{R18,R19}, we iterate the generalized DT of equations (\ref{Eq1}) and (\ref{Eq2}) and work out a formula for generation of higher-order rogue wave solutions. It is worth noting that the rogue wave solutions of this system have early been constructed in ref. \cite{R20}, using however a non-recursive Darboux transformation up to second-order. Thus, the organization of this paper is settled as follows. In section 2, we first propose, by the Darboux matrix method and the Lax-pairs, a DT for the focusing VNLSE, we then use this DT to derive, via the Taylor expansion of the solution to the Lax-pairs, the generalized DT for the focusing VNLSE. Also, we provide, through iterations of the DT, formulae for Nth-order rogue wave solutions of the focusing VNLSE. In section 3, we consider the dynamics for these solutions from the first to third-order rogue waves and show their interesting structures. The last section is devoted to a brief conclusion.

\section{Generalized Darboux transformation}
\begin{figure}
   \includegraphics[width=15cm,height=5cm]{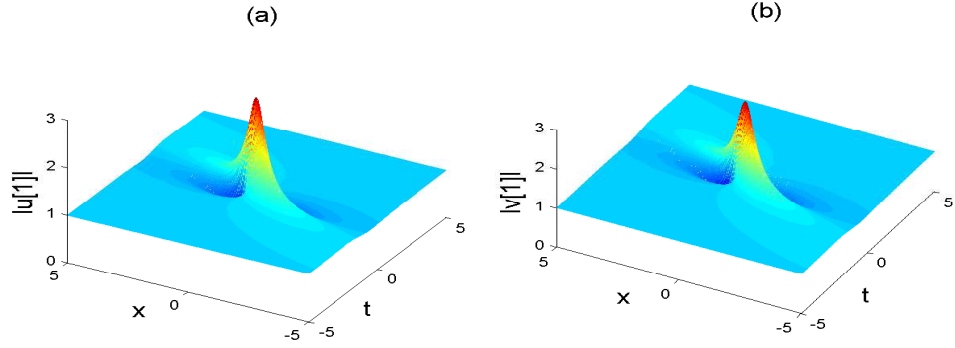}
  \caption[fundamental rogue wave]{First-order rogue waves $u[1]$ and $v[1]$ depicted with the following parameters (a) $a=1$, $b=1/4$, and (b) $a=1/4$, $b=1$, $\alpha=1$.} \label{fig1}
\end{figure}

Equations (\ref{Eq1}) and (\ref{Eq2}) can be cast into a $3\times3$ linear eigenvalue problem due to integrability \cite{R20}
  \begin{equation}\label{Eq3}
    \mathbf{R}_{x}=\mathbf{U}\mathbf{R}, \quad \mathbf{R}_{t}=\mathbf{V}\mathbf{R},
  \end{equation}
  where,
  \begin{equation}\label{Eq4}
    \mathbf{U}=\lambda\mathbf{E}+\mathbf{Q}, \quad \mathbf{V}=\frac{3}{2}\lambda^{2}\mathbf{E}+\frac{3}{2}\lambda\mathbf{Q}+\frac{i}{2}\mathbf{\sigma}_{3}(\mathbf{Q}_{x}-\mathbf{Q}^{2}),
  \end{equation}
  \begin{figure}[t]
   \includegraphics[width=15cm,height=5cm]{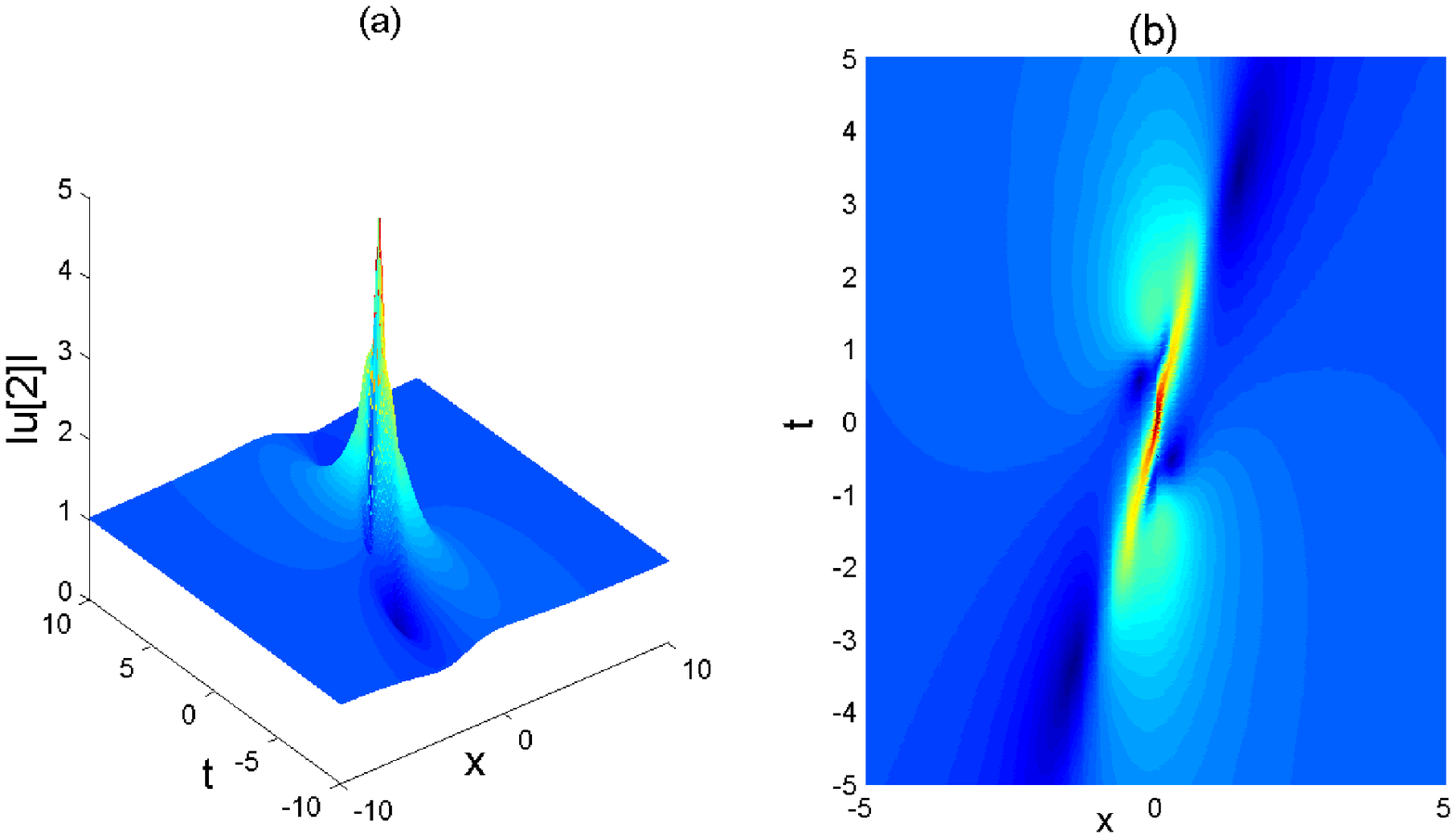}
  \caption[fundamental rogue wave]{Second-order composite rogue wave  $u[2]$ depicted with the following parameters:  $a=1$, $b=0$, $c_{1}=d_{1}=0$ and $\alpha=1$. Panel (a) represents the 3-D perspective and the panel (b) stands for the density plot of the 3-D representation.} \label{fig2}
\end{figure}
  with,
    $ \mathbf{E}=diag(-2i,i,i)$, $\mathbf{\sigma}_{3}=diag(1,-1,-1)$, $\mathbf{Q}=\left(
                                                                                         \begin{array}{ccc}
                                                                                           0 & -u & -v \\
                                                                                           u^{*} & 0 & 0 \\
                                                                                           v^{*} & 0 & 0 \\
                                                                                         \end{array}
                                                                                       \right)$.
\begin{figure}[t]
   \includegraphics[width=15cm,height=5cm]{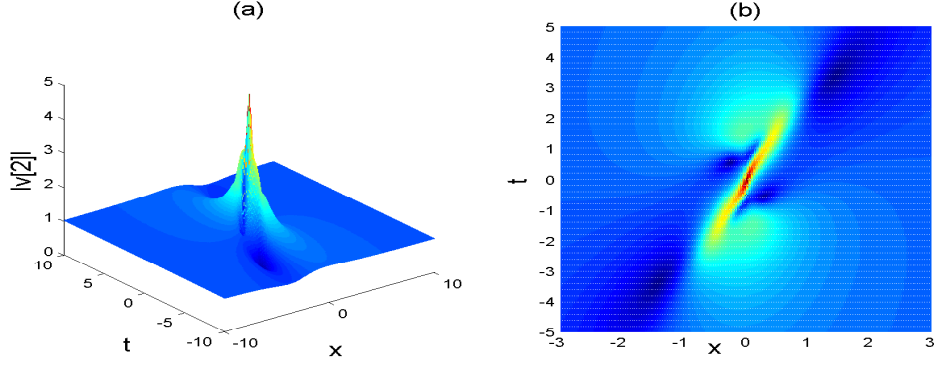}
  \caption[fundamental rogue wave]{Second-order composite rogue wave  $v[2]$ depicted with the following parameters:  $a=0$, $b=1$, $c_{1}=d_{1}=0$ and $\alpha=1$. Panel (a) represents the 3-D perspective and the panel (b) stands for the density plot of the 3-D representation. } \label{fig3}
\end{figure}

Here $\mathbf{R}=(r(x,t), s(x,t), w(x,t))^{T}$ (T means a matrix transpose), $u$ and $v$ are potentials, $\lambda$ is the spectral parameter, $u^{*}$ and $v^{*}$ denote complex conjugate of $u$ and $v$. Through direct calculation, one can directly obtain (\ref{Eq1}) and (\ref{Eq2}) by using the zero curvature equation $\mathbf{U}_{t}-\mathbf{V}_{x}+\mathbf{U}\mathbf{V}-\mathbf{V}\mathbf{U}=\mathbf{0}$.

Let $\mathbf{R}_{1}=(r_{1},s_{1},w_{1})^{T}$ be a solution of the Lax-pairs given in equation (\ref{Eq3}) with $u=u[0]$, $v=v[0]$ and $\lambda=\lambda_{1}$. The classical DT of the Ablowitz-Kaup-Newell-Segur (AKNS) spectral problem  \cite{R18} allows us to write the following formulas:
\begin{equation}\label{Eq5}
   \mathbf{R}[1]=T[1]\mathbf{R},\qquad T[1]=\lambda_{1}\mathbf{I}-H[0]\Lambda_{1}H[0]^{-1},
\end{equation}
\begin{equation}\label{Eq6}
   u[1]=u[0]+2i(\lambda-\lambda^{*})\frac{r_{1}[0]s_{1}[0]^{*}}{|r_{1}[0]|^{2}+|s_{1}[0]|^{2}+|w_{1}[0]|^{2}},
\end{equation}
\begin{equation}\label{Eq7}
     v[1]=v[0]+2i(\lambda-\lambda^{*})\frac{r_{1}[0]w_{1}[0]^{*}}{|r_{1}[0]|^{2}+|s_{1}[0]|^{2}+|w_{1}[0]|^{2}},
\end{equation}
which satisfy
\begin{equation}\label{Eq8}
    \mathbf{R}[1]_{x}=\mathbf{U}[1] \mathbf{R}[1], \qquad \mathbf{R}[1]_{t}=\mathbf{V}[1]\mathbf{R}[1],
\end{equation}
where $\mathbf{R}_{1}[0]=T[0]\mathbf{R}_{1}$, $r_{1}[0]=r_{1}$, $s_{1}[0]=r_{1}$, $r_{1}[0]=r_{1}$,
   $ \mathbf{I}=\left(
                 \begin{array}{ccc}
                   1 & 0 & 0 \\
                   0 & 1 & 0 \\
                   0 & 0 & 1 \\
                 \end{array}
               \right),$ $ H[0]=\left(
                                      \begin{array}{ccc}
                                        r_{1}[0] &s_{1}[0]^{*}  &  w_{1}[0]^{*} \\
                                        s_{1}[0] &-r_{1}[0]^{*} & 0 \\
                                         w_{1}[0]& 0 & -r_{1}[0]^{*} \\
                                      \end{array}
                                    \right)$,
\begin{figure}[t]
   \includegraphics[width=15cm,height=5cm]{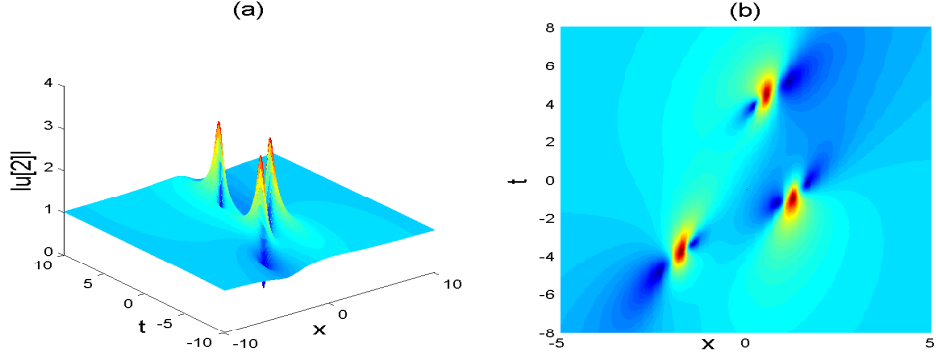}
  \caption[fundamental rogue wave]{Second-order  rogue wave  $u[2]$ depicted with the following parameters:  $a=1$, $b=0$, $c_{1}=d_{1}=25$ and $\alpha=1$. Panel (a) represents the 3-D perspective and the panel (b) stands for the density plot of the 3-D representation. } \label{fig4}
\end{figure}

and
$    \Lambda_{1}=\left(\begin{array}{ccc}
                      \lambda_{1} & 0 & 0 \\
                      0 & \lambda_{1}^{*} & 0 \\
                      0 & 0 & \lambda_{1}^{*} \\
                      \end{array}
                      \right)$.
The quantities $\mathbf{U}[1]$ and $\mathbf{V}[1]$ have the same form as $\mathbf{U}$ and $\mathbf{V}$ except that the old potentials $u$ and $v$ are replaced by the new ones $u[1]$ and $v[1]$. The quantity $T$ stands for the Darboux matrix.

If N distinct basic solutions $\mathbf{R}_{k}=(r_{k},s_{k},w_{k})^{T}$ $(k=1,2,3,...,N)$ of the Lax-pairs expressed by equation (\ref{Eq3}) at $\lambda=\lambda_{k}$ $(k=1,2,3,...,N)$ are given, the DT can be repeated N times. Then, the Nth-step DT for the VNLSE  (\ref{Eq1}) and (\ref{Eq2}) is
\begin{eqnarray}\label{Eq9}
\mathbf{R}[N]=T[N]T[N-1]...T[1]\mathbf{R},\quad T[k]=\lambda\mathbf{I}-H[k-1]\Lambda_{k}H[k-1]^{-1},
\end{eqnarray}
\begin{eqnarray}\label{Eq10}
    u[N]&=&u[N-1]+2i(\lambda-\lambda^{*})\frac{r_{N}[N-1]s_{N}[N-1]^{*}}{|r_{N}[N-1]|^{2}+|s_{N}[N-1]|^{2}+|w_{N}[N-1]|^{2}},
\end{eqnarray}
\begin{eqnarray}\label{Eq11}
    v[N]&=&v[N-1]+2i(\lambda-\lambda^{*})\frac{r_{N}[N-1]w_{N}[N-1]^{*}}{|r_{N}[N-1]|^{2}+|s_{N}[N-1]|^{2}+|w_{N}[N-1]|^{2}},
\end{eqnarray}

where

$                 H[k-1]=\left(            \begin{array}{ccc}
                                        r_{k}[k-1] &s_{k}[k-1]^{*}  &  w_{k}[k-1]^{*} \\
                                        s_{k}[k-1] &-r_{k}[k-1]^{*} & 0 \\
                                         w_{k}[k-1]& 0 & -r_{k}[k-1]^{*} \\
                                      \end{array}
                                    \right)$,
$     \Lambda_{k}=\left(\begin{array}{ccc}
                      \lambda_{k} & 0 & 0 \\
                      0 & \lambda_{k}^{*} & 0 \\
                      0 & 0 & \lambda_{k}^{*} \\
                       \end{array}
                       \right)$,

$\mathbf{R}_{k}[k-1]=(r_{k}[k-1],s_{k}[k-1],w_{k}[k-1])^{T}=\mathbf{R}_{k}[k-1]$ and $\mathbf{R}[k-1]=T[k-1]T[k-2]...T[1]\mathbf{R}$.
 \begin{figure}[t]
   \includegraphics[width=15cm,height=5cm]{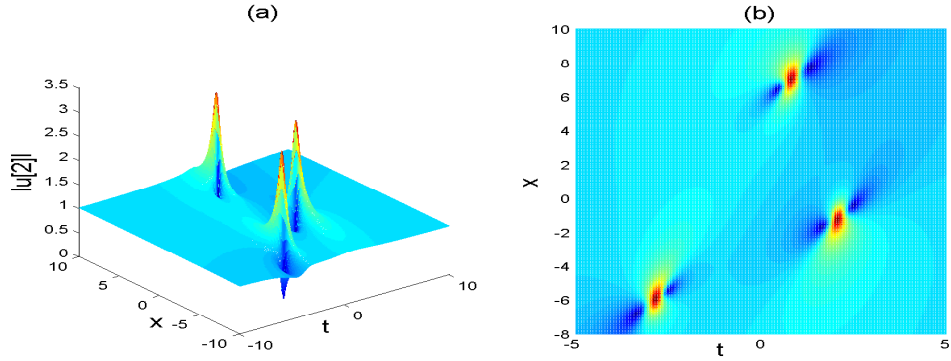}
  \caption[fundamental rogue wave]{Second-order  rogue wave  $u[2]$ depicted with the following parameters:  $a=1$, $b=0$, $c_{1}=d_{1}=100$ and $\alpha=1$. Panel (a) represents the 3-D perspective and the panel (b) stands for the density plot of the 3-D representation. } \label{fig5}
\end{figure}

  According to the above classical DT, we derive the generalized DT for the system of equations (\ref{Eq1}) and (\ref{Eq2}). We start with the assumption that
  \begin{equation}\label{Eq12}
   R_{1}=R_{1}(\lambda_{1}+\epsilon),
  \end{equation}
  is a particular solution of the Lax-pairs of equation (\ref{Eq3}). The constant $\epsilon$ being a small parameter. Expanding $R_{1}$ in a Taylor series gives
  \begin{equation}\label{Eq13}
    R_{1}=R_{1}^{[0]}+R_{1}^{[1]}\epsilon+R_{1}^{[2]}\epsilon^{2}+...+R_{1}^{[N]}\epsilon^{N}+...,
  \end{equation}
  where $R_{1}^{[k]}=\frac{1}{k!}\frac{\partial^{k}}{\partial \epsilon^{k}}R_{1}|_{\epsilon=0}$ $(k=1,2,3,...)$.\\

  1. \emph{The first-step of the method.}

  From the above assumption, it is easy to find that $R_{1}^{[0]}$ is a solution for the Lax-pairs of equation (\ref{Eq3}) with $u=u[0]$ and $v=v[0]$ at $\lambda=\lambda_{1}$. The first-step DT for the system (\ref{Eq1}) and (\ref{Eq2}) is expressed as
   \begin{equation}\label{Eq14}
    \mathbf{R}[1]=T[1]\mathbf{R},\qquad T[1]=\lambda_{1}\mathbf{I}-H[0]\Lambda_{1}H[0]^{-1},
   \end{equation}
   \begin{equation}\label{Eq15}
     u[1]=u[0]+2i(\lambda-\lambda^{*})\frac{r_{1}[0]s_{1}[0]^{*}}{|r_{1}[0]|^{2}+|s_{1}[0]|^{2}+|w_{1}[0]|^{2}},
   \end{equation}
      \begin{equation}\label{Eq16}
     v[1]=v[0]+2i(\lambda-\lambda^{*})\frac{r_{1}[0]w_{1}[0]^{*}}{|r_{1}[0]|^{2}+|s_{1}[0]|^{2}+|w_{1}[0]|^{2}},
   \end{equation}
    where

    $\mathbf{R}_{1}[0]=T[0]\mathbf{R}_{1}$, $r_{1}[0]=r_{1}$, $s_{1}[0]=r_{1}$, $r_{1}[0]=r_{1}$,
    \begin{figure}[t]
   \includegraphics[width=15cm,height=5cm]{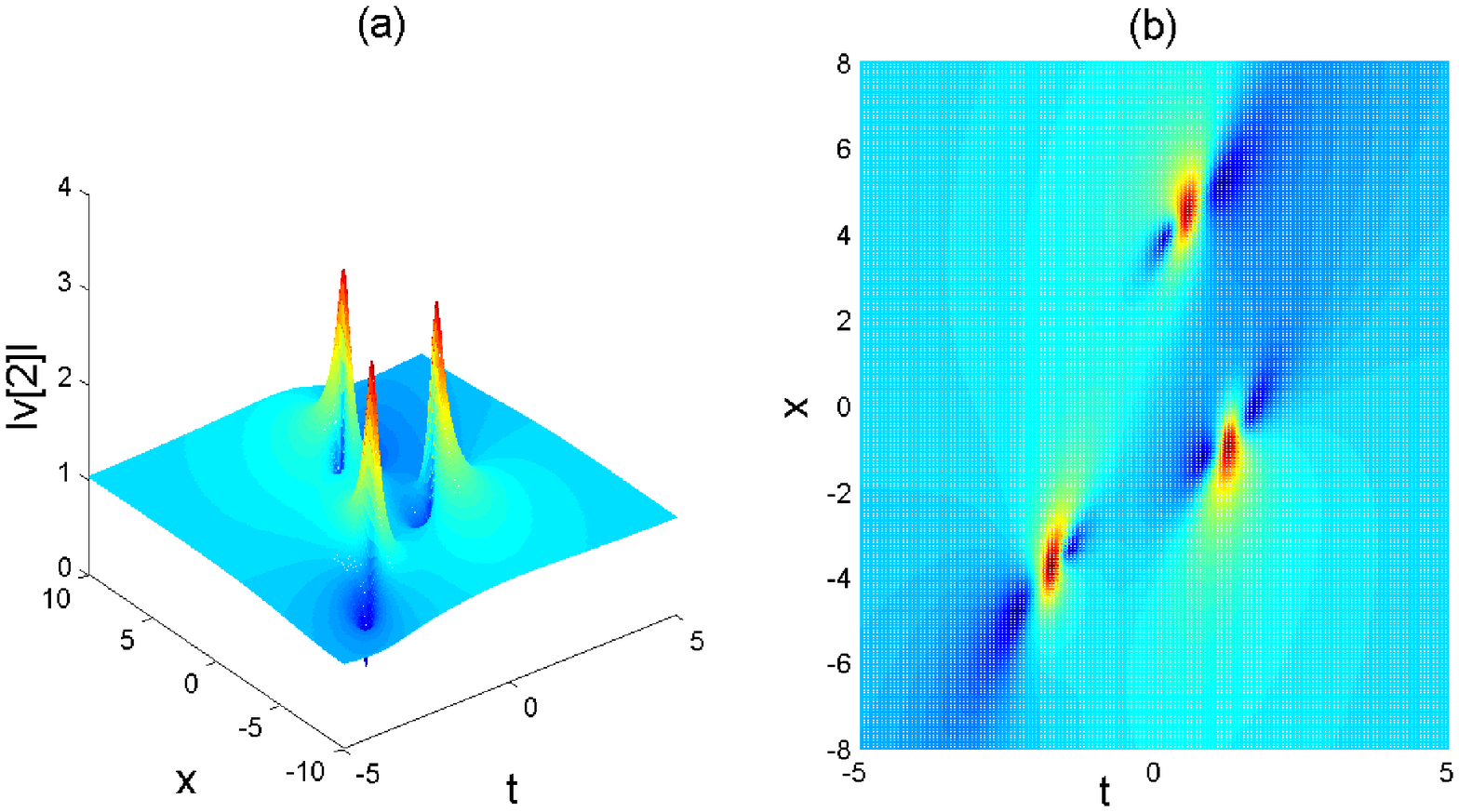}
  \caption[fundamental rogue wave]{Second-order  rogue wave  $v[2]$ depicted with the following parameters:  $a=0$, $b=1$, $c_{1}=d_{1}=25$ and $\alpha=1$. Panel (a) represents the 3-D perspective and the panel (b) stands for the density plot of the 3-D representation. } \label{fig6}
\end{figure}
$    \mathbf{I}=\left(
                 \begin{array}{ccc}
                   1 & 0 & 0 \\
                   0 & 1 & 0 \\
                   0 & 0 & 1 \\
                 \end{array}
               \right)$, $ H[0]=\left(
                                      \begin{array}{ccc}
                                        r_{1}[0] &s_{1}[0]^{*}  &  w_{1}[0]^{*} \\
                                        s_{1}[0] &-r_{1}[0]^{*} & 0 \\
                                         w_{1}[0]& 0 & -r_{1}[0]^{*} \\
                                      \end{array}
                                    \right)$,

and
$    \Lambda_{1}=\left(\begin{array}{ccc}
                      \lambda_{1} & 0 & 0 \\
                      0 & \lambda_{1}^{*} & 0 \\
                      0 & 0 & \lambda_{1}^{*} \\
                      \end{array}
                      \right)$.

2. \emph{The second-step of the method}.

With
\begin{eqnarray}\label{Eq17}
\lim_{\epsilon \rightarrow 0}\frac{T[1]|_{\lambda=\lambda_{1}+\epsilon}R_{1}}{\epsilon}=\lim_{\epsilon \rightarrow 0}\frac{(\epsilon+T_{1}[1])R_{1}}{\epsilon}=R_{1}^{[0]}+T_{1}[1]R_{1}^{[1]}\equiv R_{1}[1],
 \end{eqnarray}

 we find a solution to the Lax-pairs  of equation (\ref{Eq3}) with $u[2]$ and $v[2]$ and $\lambda=\lambda_{1}$. This allows us to go to the second step DT, namely,
 \begin{figure}[t]
   \includegraphics[width=15cm,height=5cm]{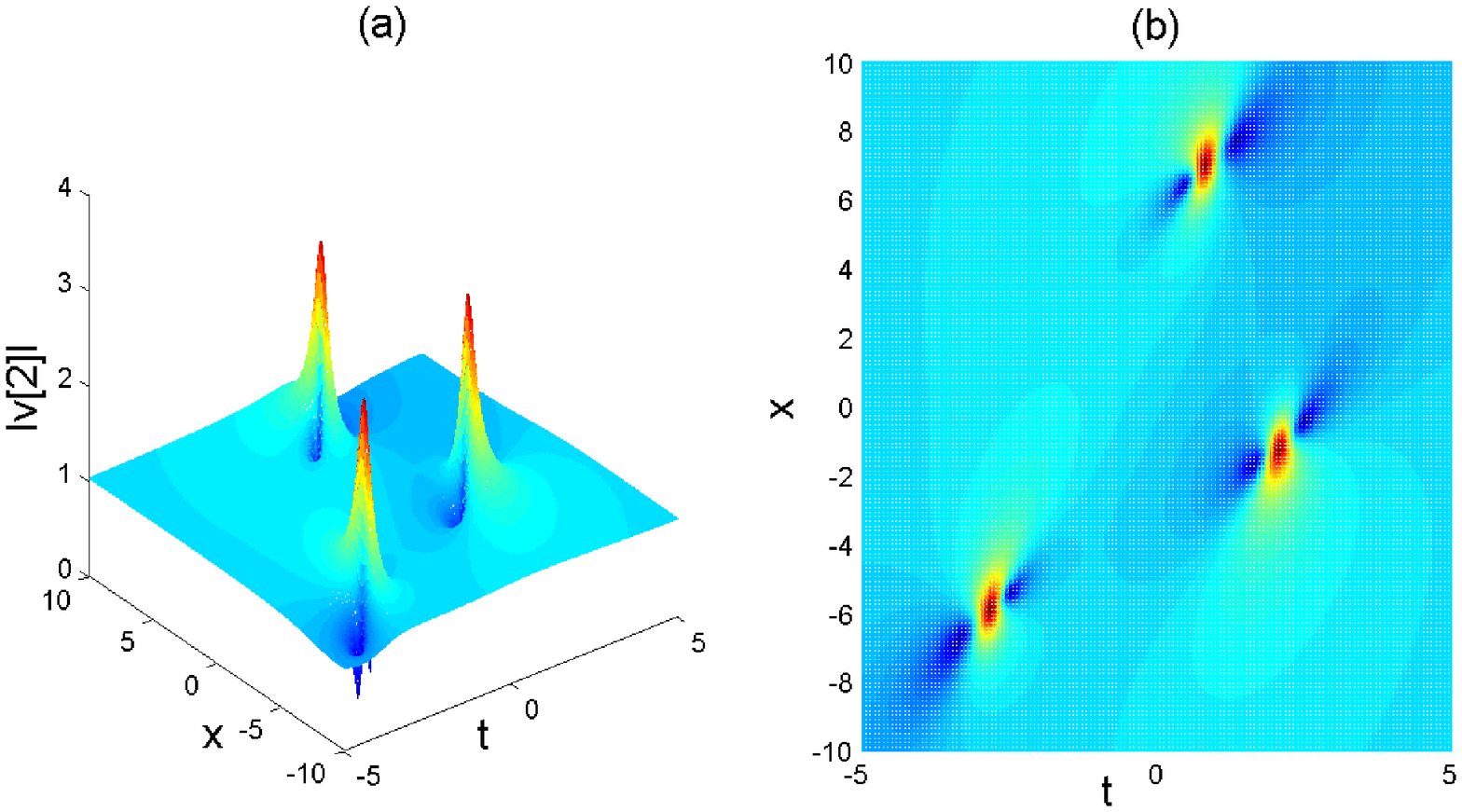}
  \caption[fundamental rogue wave]{Second-order  rogue wave  $v[2]$ depicted with the following parameters:  $a=0$, $b=1$, $c_{1}=d_{1}=100$ and $\alpha=1$. Panel (a) represents the 3-D perspective and the panel (b) stands for the density plot of the 3-D representation. } \label{fig7}
\end{figure}
 \begin{equation}\label{Eq18}
 R[2]=T[2]T[1]R,\quad T[2]=\lambda_{1}\mathbf{I}-H[1]\Lambda_{2}H[1]^{-1},
  \end{equation}
    \begin{equation}\label{Eq19}
    u[2]=u[1]+2i(\lambda-\lambda^{*})\frac{r_{1}[1]s_{1}[1]^{*}}{|r_{1}[1]|^{2}+|s_{1}[1]|^{2}+|w_{1}[1]|^{2}},
 \end{equation}
 \begin{equation}\label{Eq20}
     v[2]=v[1]+2i(\lambda-\lambda^{*})\frac{r_{1}[1]w_{1}[1]^{*}}{|r_{1}[1]|^{2}+|s_{1}[1]|^{2}+|w_{1}[1]|^{2}},
 \end{equation}
 where $(r_{1}[1],s_{1}[1],w_{1}[1])^{T}=R_{1}[1]$,
$    \mathbf{I}=\left(
                 \begin{array}{ccc}
                   1 & 0 & 0 \\
                   0 & 1 & 0 \\
                   0 & 0 & 1 \\
                 \end{array}
               \right), $ $ H[1]=\left(
                                      \begin{array}{ccc}
                                        r_{1}[1] &s_{1}[1]^{*}  &  w_{1}[1]^{*} \\
                                        s_{1}[1] &-r_{1}[1]^{*} & 0 \\
                                         w_{1}[1]& 0 & -r_{1}[1]^{*} \\
                                      \end{array}
                                    \right)$,

  and
 $    \Lambda_{2}=\left(\begin{array}{ccc}
                             \lambda_{1} & 0 & 0 \\
                             0 & \lambda_{1}^{*} & 0 \\
                             0 & 0 & \lambda_{1}^{*} \\
                             \end{array}
                             \right)$.

 3. \emph{The third-step of the method}.

\begin{figure}[t]
   \includegraphics[width=15cm,height=5cm]{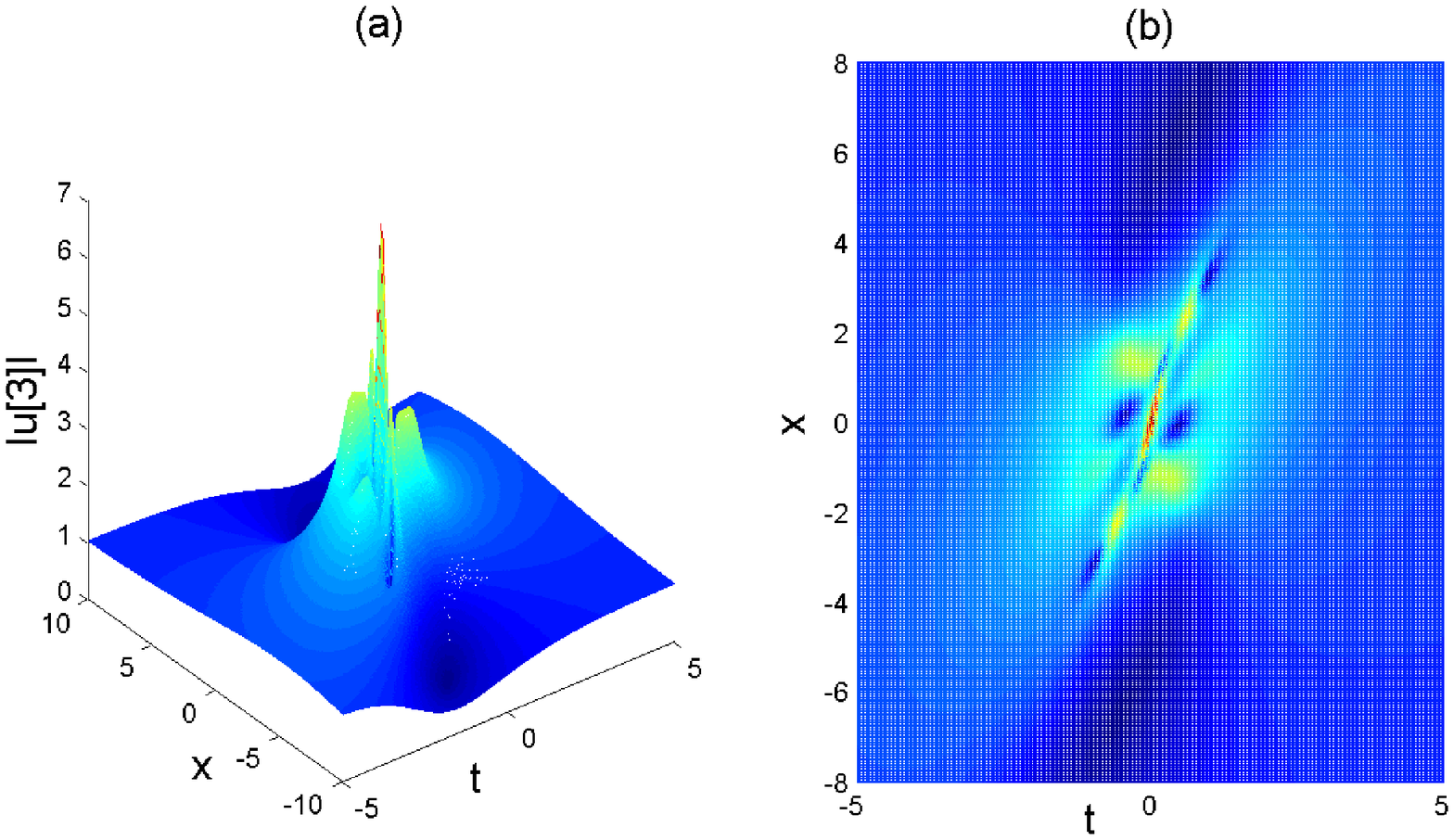}
  \caption[fundamental rogue wave]{Third-order composite  rogue wave  $u[3]$ depicted with the following parameters:  $a=1$, $b=0$, $c_{1}=d_{1}=c_{2}=d_{2}=0$ and $\alpha=1$. Panel (a) represents the 3-D perspective and the panel (b) stands for the density plot of the 3-D representation. } \label{fig8}
\end{figure}

   Similarly, the following limit
   \begin{eqnarray}\label{Eq21}
   \lim_{\epsilon \rightarrow 0}\frac{[T[2]T[1]]|_{\lambda =\lambda_{1}+\epsilon}R_{1}}{\epsilon^{2}} &=&  \lim_{\epsilon \rightarrow 0}\frac{(\epsilon+T_{1}[2])(\epsilon+T_{1}[1])R_{1}}{\epsilon^{2}}\nonumber\\
  && =R_{1}^{[0]}+(T_{1}[2]+T_{1}[1])R_{1}^{[1]}+T_{1}[2]T_{1}[1]R_{1}^{[2]}\equiv R_{1}[2],
   \end{eqnarray}
 provides us with a non-trivial solution for the Lax-pairs of equation (\ref{Eq3}) with $u[3]$ , $v[3]$ and $\lambda=\lambda_{1}$. Then, the third-step generalized DT can be given as follows
  \begin{equation}\label{Eq22}
 R[3]=T[3]T[2]T[1]R,\quad T[3]=\lambda_{1}\mathbf{I}-H[2]\Lambda_{3}H[2]^{-1},
  \end{equation}
\begin{equation}\label{Eq23}
    u[3]=u[2]+2i(\lambda-\lambda^{*})\frac{r_{1}[2]s_{1}[2]^{*}}{|r_{1}[2]|^{2}+|s_{1}[2]|^{2}+|w_{1}[2]|^{2}},
 \end{equation}
 \begin{equation}\label{Eq24}
     v[3]=v[2]+2i(\lambda-\lambda^{*})\frac{r_{1}[2]w_{1}[2]^{*}}{|r_{1}[2]|^{2}+|s_{1}[2]|^{2}+|w_{1}[2]|^{2}},
 \end{equation}

 where $(r_{1}[2],s_{1}[2],w_{1}[2])^{T}=R_{1}[2]$,
 $    \mathbf{I}=\left(
                 \begin{array}{ccc}
                   1 & 0 & 0 \\
                   0 & 1 & 0 \\
                   0 & 0 & 1 \\
                 \end{array}
               \right)$, $ H[2]=\left(
                                      \begin{array}{ccc}
                                        r_{1}[2] &s_{1}[2]^{*}  &  w_{1}[2]^{*} \\
                                        s_{1}[2] &-r_{1}[2]^{*} & 0 \\
                                         w_{1}[2]& 0 & -r_{1}[2]^{*} \\
                                      \end{array}
                                    \right)$,

  \begin{figure}
   \includegraphics[width=15cm,height=5cm]{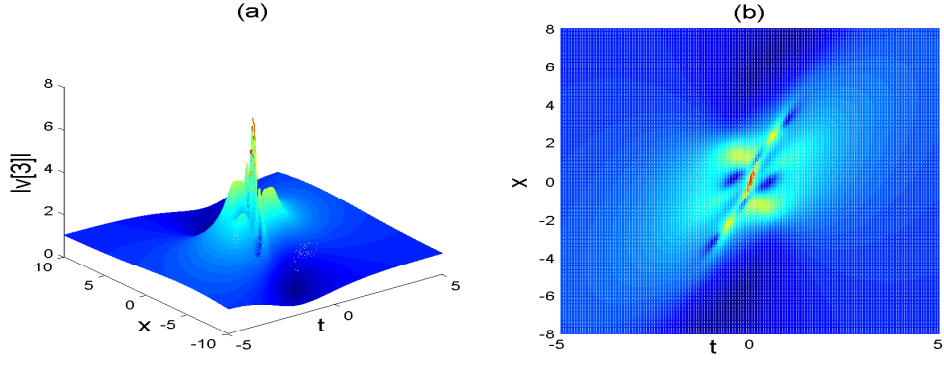}
  \caption[fundamental rogue wave]{Third-order composite  rogue wave  $v[3]$ depicted with the following parameters:  $a=0$, $b=1$, $c_{1}=d_{1}=c_{2}=d_{2}=0$ and $\alpha=1$. Panel (a) represents the 3-D perspective and the panel (b) stands for the density plot of the 3-D representation. } \label{fig9}
\end{figure}
   and
  $     \Lambda_{3}=\left( \begin{array}{ccc}
                      \lambda_{1} & 0 & 0 \\
                       0 & \lambda_{1}^{*} & 0 \\
                       0 & 0 & \lambda_{1}^{*} \\
                       \end{array}
                       \right)$.

 4. \emph{The Nth-step generalized DT}.

 Continuing the above process and combining all the Darboux matrices, an Nth-step generalized DT is considered as follows:

\begin{eqnarray}\label{Eq25}
  R_{1}[N-1]=R_{1}^{[0]}+\sum_{k=1}^{N-1}T_{1}[k]R_{1}^{[1]}+\sum_{k=1}^{N-1}\sum_{l=1}^{k-1}T_{1}[k]T_{1}[l]R_{1}^{[2]}+... +T_{1}[N-1]T_{1}[N-2]...T_{1}[1]R_{1}^{[N-1]},
\end{eqnarray}

 \begin{eqnarray}\label{Eq26}
 R[N]=T[N]T[N-1]...T[1]R,\quad T[k]=\lambda_{1}\mathbf{I}-H[k-1]\Lambda_{k}H[k-1]^{-1},
  \end{eqnarray}

 \begin{eqnarray}\label{Eq27}
   u[N] &=& u[N-1]+2i(\lambda-\lambda^{*})\frac{r_{1}[N-1]s_{1}[N-1]^{*}}{|r_{1}[N-1]|^{2}+|s_{1}[N-1]|^{2}+|w_{1}[N-1]|^{2}},
 \end{eqnarray}
 \begin{eqnarray}\label{Eq28}
     v[N]&=& v[N-1]+2i(\lambda-\lambda^{*})\frac{r_{1}[N-1]w_{1}[N-1]^{*}}{|r_{1}[N-1]|^{2}+|s_{1}[N-1]|^{2}+|w_{1}[N-1]|^{2}},
 \end{eqnarray}

where $(r_{1}[N-1],s_{1}[N-1],w_{1}[N-1])^{T}=R_{1}[N-1]$
$     H[k-1]=\left(
                  \begin{array}{ccc}
                  r_{1}[k-1] &s_{1}[k-1]^{*}  &  w_{1}[k-1]^{*} \\
                   s_{1}[k-1] &-r_{1}[k-1]^{*} & 0 \\
                   w_{1}[k-1]& 0 & -r_{1}[k-1]^{*} \\
                    \end{array}
                       \right)$,

 $     \Lambda_{k}=\left(
                        \begin{array}{ccc}
                        \lambda_{1} & 0 & 0 \\
                        0 & \lambda_{1}^{*} & 0 \\
                        0 & 0 & \lambda_{1}^{*} \\
                        \end{array}
                        \right)$,            and                           $ \mathbf{I}=\left(
                                                        \begin{array}{ccc}
                                                        1 & 0 & 0 \\
                                                        0 & 1 & 0 \\
                                                        0 & 0 & 1 \\
                                                        \end{array}
                                                            \right)$.
  \begin{figure}[t]
   \includegraphics[width=15cm,height=5cm]{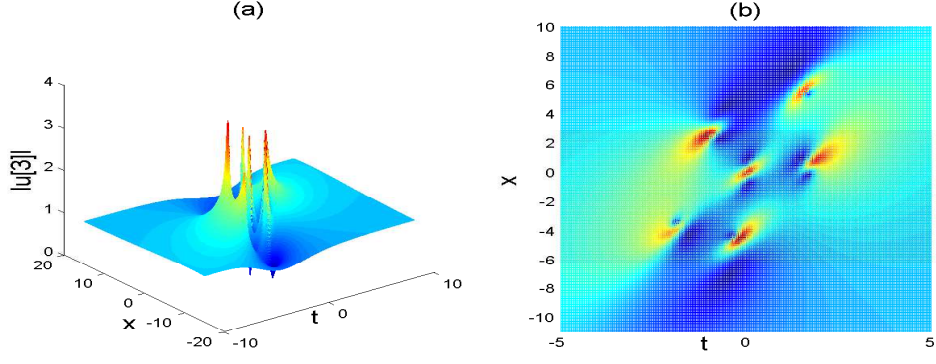}
  \caption[fundamental rogue wave]{Third-order  rogue wave  $u[3]$ depicted with the following parameters:  $a=1$, $b=0$, $c_{1}=d_{1}=0$, $c_{2}=d_{2}=100$ and $\alpha=1$.  Panel (a) represents the 3-D perspective and the panel (b) stands for the density plot of the 3-D representation} \label{fig10}
\end{figure}

 The formulae given by equations (\ref{Eq26})-(\ref{Eq28}) are a recursive formula of the Nth-order generalized DT for the Manakov system (\ref{Eq1}) and (\ref{Eq2}). Although it is possible to give the $(3N)\times(3N)$ determinant representation by using the so-called crum theorem \cite{R21}. But we prefer to use a recursive formula, because it is easy to construct higher-order rogue wave solutions with the computer. Some interesting higher-order rogue wave solutions are obtained for the Manakov system in the following section.

\section{Rogue wave solutions}

In order to obtain the rogue wave solution, we start with the following seed solutions of the system (\ref{Eq1}) and (\ref{Eq2}) as
 \begin{equation}\label{Eq29}
   u[0]=a e^{i \beta t}, \quad v[0]=b e ^{i \beta t},
 \end{equation}
 with $\beta=a^{2}+b^{2}$, $a$ and $b$ are real constants. Then the  basic solution for the Lax-pairs of equation (\ref{Eq3}) with $u[0]$, $v[0]$   and  $\lambda$   holds
  \begin{equation}\label{Eq30}
    R_{1}=\left(
            \begin{array}{c}
              (m_{1} e^{\eta_{1}+\eta_{2}}-m_{2} e^{\eta_{1}-\eta_{2}} ) e^{\frac{i \beta t}{2}}  \\
             \tau_{1}(m_{2} e^{\eta_{1}+\eta_{2}}-m_{1} e ^{\eta_{1}-\eta_{2}} ) e^{-\frac{i \beta t}{2}}  \\
             \tau_{2}(m_{2} e^{\eta_{1}+\eta_{2}}-m_{1} e^{\eta_{1}-\eta_{2}} )e^{-\frac{i \beta t}{2}}  \\
            \end{array}
          \right),
 \end{equation}
 where
     $m_{1}=\left(\frac{\lambda-\sqrt{\beta+\lambda^{2}}}{\lambda^{2}+\beta}\right)^{\frac{1}{2}},$  $  m_{2}=\left(\frac{\lambda+\sqrt{\beta+\lambda^{2}}}{\lambda^{2}+\beta}\right)^{\frac{1}{2}},$
 $     \eta_{1}=2i\alpha(\frac{1}{\alpha}x+\lambda(\lambda+2)t)$, $ \tau_{1}=\frac{a}{\sqrt{\beta}} $, $\tau_{2}=\frac{b}{\sqrt{\beta}}$,

        $ \eta_{2}=i\alpha \sqrt{\beta+\lambda^{2}}\left(\frac{1}{\alpha}x+(\lambda(\lambda+2)-\frac{\beta}{2})t+\sum_{j=1}^{N}(c_{j}+id_{j})\varepsilon^{2j}\right).$

 Here the constant $\varepsilon$ is a small parameter.

 Let $\lambda=i\sqrt{\beta}(1+\varepsilon^{2})$, expanding the vector function $R_{1}(\varepsilon)$ at $\varepsilon=0$, we obtain
   \begin{equation}\label{Eq31}
   R_{1}(\varepsilon)=R_{1}^{[0]}+R_{1}^{[1]}\varepsilon^{2}+R_{1}^{[2]}\varepsilon^{4}+...,
 \end{equation}
where,

   $   R_{1}^{[0]}=\left(
                 \begin{array}{c}
                   r_{1}^{0} \\
                  s_{1}^{0} \\
                  w_{1}^{0} \\
                 \end{array}
               \right),$ $  R_{1}^{[1]}=\left(
                                               \begin{array}{c}
                                                 r_{1}^{1} \\
                                                 s_{1}^{1} \\
                                                 w_{1}^{1} \\
                                                 \end{array}
                                                    \right)$, $ R_{1}^{[2]}=\left(
                                                                                   \begin{array}{c}
                                                                                   r_{1}^{2} \\
                                                                                    s_{1}^{2} \\
                                                                                    w_{1}^{2} \\
                                                                                    \end{array}
                                                                                         \right)$...,
    and $(r_{1}^{[i-1]},s_{1}^{[i-1]},w_{1}^{[i-1]})$ (i=1,2,3) are given in appendix.

  It is clear that $R_{1}^{[0]}$ is a solution of the Lax pairs (\ref{Eq3}) at $u[0]=a e^{i \beta t}$ , $v[0]=b e^{i \beta t}$ and $\lambda=i\sqrt{\beta}(1+\varepsilon^{2})$. Hence from the formulae (\ref{Eq15}) and (\ref{Eq16}), we arrive at
  \begin{equation}\label{Eq32}
    u[1]=a e^{i\beta t}\left(1+\frac{F_{1}+iH_{1}}{D_{1}}\right), \quad v[1]= b e^{i\beta t}\left(1+\frac{F_{1}+iH_{1}}{D_{1}}\right),
  \end{equation}
  where

 $ F_{1}=-4 \sqrt{\beta} \alpha\left(4\beta x^{2}-12\alpha \beta^{2} t x +\alpha^{2}\beta^{2}(9\beta+16)t^{2}-1\right)$,

  $ H_{1}=32\alpha \sqrt{\beta} t,$ $ D_{1}=\frac{2}{\beta}\left(4\beta x^{2}-12\alpha \beta^{2} t x +\alpha^{2}\beta^{2}(9\beta+16)t^{2}+1\right).  $

  The solutions $u[1]$ and $v[1]$ here stand for the vector generalized  first-order rogue wave solutions for the Manakov system (\ref{Eq1}) and (\ref{Eq2}). It is important to remark that, $u[1]$ and $v[1]$ are merely proportional. These solutions are depicted in figure 1. It is also important to note that the solutions obtained here for the Manakov system resemble those obtained in ref \cite{R20}.
      \begin{figure}[t]
   \includegraphics[width=15cm,height=5cm]{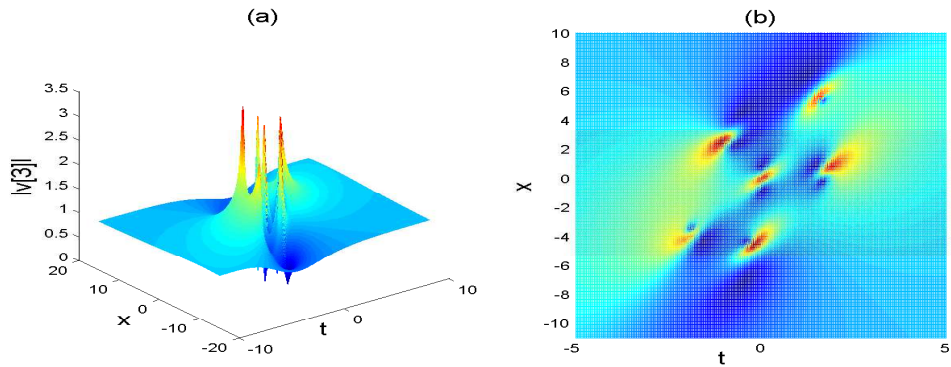}
  \caption[fundamental rogue wave]{Third-order rogue wave $v[3]$ depicted with the following parameters:   $a=0$, $b=1$, $c_{1}=d_{1}=0$, $c_{2}=d_{2}=100$ and $\alpha=1$. Panel (a) represents the 3-D perspective and the panel (b) stands for the density plot of the 3-D representation } \label{fig11}
\end{figure}

     Then, using the matrices $R_{1}^{[0]}$ and $R_{1}^{[1]}$ given in appendix, and substituting them into the expression of equation (\ref{Eq17}), we obtain the matrix $R_{1}[1]$ which elements are also given in appendix. The matrix elements $r_{1}[1]$, $s_{1}[1]$ and $w_{1}[1]$ are then substituted into equations (\ref{Eq19}) and (\ref{Eq20}), which give rise to the second-order vector generalization rogue wave solutions of the Manakov system (\ref{Eq1}) and (\ref{Eq2}).

      The  second-order  rogue wave solution possesses two free parameters $c_{1}$ and $d_{1}$. In general, the second-order rogue solution is composed of three first-order rogue waves.  For the case where $c_{1}=d_{1}=0$, the rogue wave are crowded round the origin (0,0) and the maximum of $u[2]$  and $v[2]$ is 5. For this case we have rogue wave composite (see figures \ref{fig2} and \ref{fig3} ), which resemble those obtained in Ref \cite{R20}. When we increase the value of $|c_{1}|$ and $|d_{1}|$ we observe that three first-order rogue waves are scattered in all direction (see figures \ref{fig4}-\ref{fig7}).

  Iterating formulae (\ref{Eq25})-(\ref{Eq28}) three times (with $N=3$), we obtain the third-order rogue wave solutions $u[3]$ and $v[3]$. We omit presenting analytical expressions  since they are rather cumbersome to be write down here. By means of computer, we give the pictures of these third-order rogue wave solutions in figures \ref{fig9} and \ref{fig10}. We know that the third-order rogue wave possesses four free parameters $c_{1}$, $d_{1}$, $c_{2}$ and $d_{2}$. The third-order rogue wave solution is composed of six first-order rogue waves. For the case where $c_{1}=d_{1}=c_{2}=d_{2}=0$, we can observe a composite third-order rogue wave solution and the maximum value of $u[3]$ and $v[3]$ is 7 (see figures \ref{fig8} and \ref{fig9}).

When $c_{1}=d_{1}=0$, $c_{2}=d_{2}=100$, the six first-order rogue wave array a  pentagon; among the six first-order rogue waves, one sits in the center and the rest are located on the vertices of the pentagon (see figures \ref{fig10} and \ref{fig11})

\begin{figure}[t]
   \includegraphics[width=15cm,height=5cm]{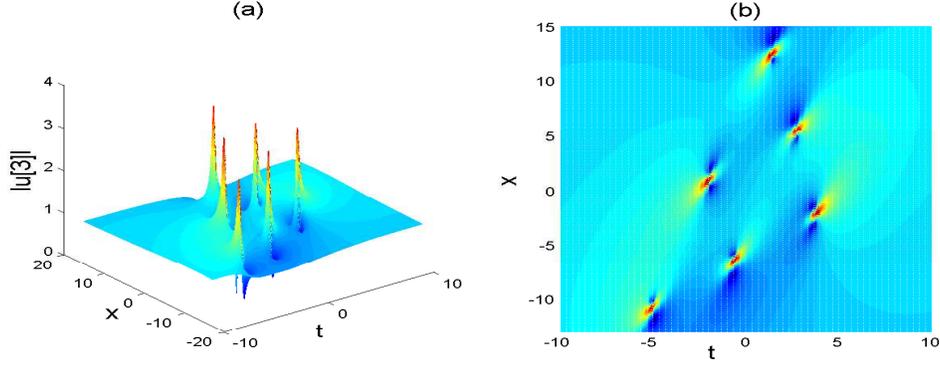}
  \caption[fundamental rogue wave]{Third-order rogue wave $u[3]$ depicted with the following parameters:   $a=1$, $b=0$, $c_{1}=d_{1}=c_{2}=d_{2}=100$ and $\alpha=1$. Panel (a) represents the 3-D perspective and the panel (b) stands for the density plot of the 3-D representation} \label{fig12}
\end{figure}

\begin{figure}[t]
   \includegraphics[width=15cm,height=5cm]{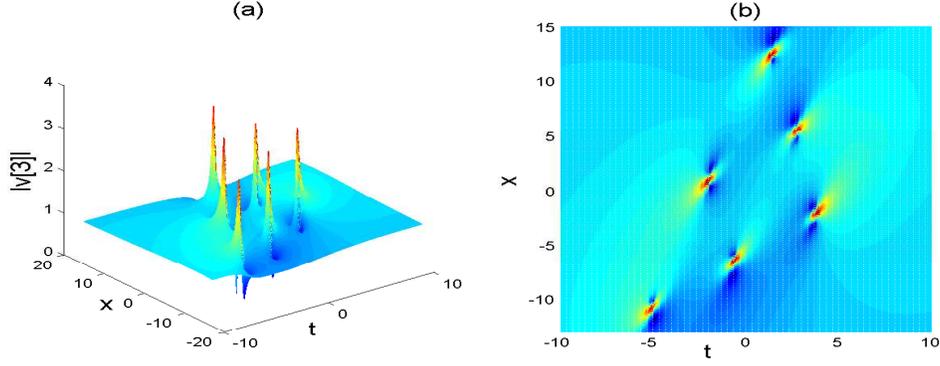}
  \caption[fundamental rogue wave]{Third-order rogue wave $v[3]$ depicted with the following parameters:   $a=0$, $b=1$, $c_{1}=d_{1}=c_{2}=d_{2}=100$ and $\alpha=1$. Panel (a) represents the 3-D perspective and the panel (b) stands for the density plot of the 3-D representation } \label{fig13}
\end{figure}

 When $c_{1}=d_{1}=100$, $c_{2}=d_{2}=100$, the corresponding third-order rogue wave is composed of six first-order rogue waves as well, which array a triangle ( see figures \ref{fig12} and \ref{fig13}).

 \section{Conclusion}

 In  this  paper,  we presented in detail a procedure  of  the construction  of  a  generalized  DT  for  the  Manakov system.
 This construction is divided into two steps. First, a brief introduction of the DT for the system (\ref{Eq1}) and (\ref{Eq2}) is  given  by  the  Darboux  matrix  method.  Then,  a  detailed derivation  of  the  generalized  DT  for system  (\ref{Eq1}) and (\ref{Eq2}) is discussed through the Taylor expansion and a limit procedure. The generalized DT allows us to calculate the higher-order rogue wave solutions for the Manakov system (\ref{Eq1}) and (\ref{Eq2}) in a unified way. In particular, some higher-order rogue wave solutions for the Manakov system are constructed by means of the generalized DT with  seed solutions $ u[0] = e^{i\beta t}$ and $v[0] = e^{i\beta t}$.

The DT, which is originally the frame work of Darboux \cite{R14}, is a powerful method for constructing solutions for integrable systems. In view to obtain rogue wave solutions for the nonlinear wave equations, the original DT is not applicable directly. So Matveev \cite{R15} introduced the so-called generalized DT and the positon solutions were calculated for the KdV equation \cite{R15}. Then, Guo et al. \cite{R13} re-examined Matveev's generalized DT and proposed a new approach to derive the generalized DT for the KdV and the NLS equations. The Guo et al \cite{R13}'s approach has been used in this work to derive higher-order rogue wave for the Manakov system (\ref{Eq1}) and (\ref{Eq2}).

 The first-order  vector rogue solutions $u[1]$ and $v[1]$ are merely proportional and it is important to point out that their maximum is three times more than their each average crest. The profiles of the higher-order rogue wave solutions for the Manakov system depend on the values of their free parameters $c_{j}$ and  $d_{j}$ ($j=1,2,3,...$). The second-order vector rogue wave solutions $u[2]$ and $v[2]$  have their maximum five times more than their each average crest and are crowded round the origin $(0,0)$ for the case where $c_{1}=d_{1}=0$. For the case $c_{1}=d_{1}=100$, we obtain three fundamental rogue waves scattered in all direction arraying a regular triangle and each fundamental rogue wave sits on a vertices. The third-order rogue waves $u[3]$ and $v[3]$ illustrated in this work have their maximum seven times more than each average crest and are crowded round the origin $(0,0)$ for the case where $c_{1}=d_{1}=c_{2}=d_{2}=0$. For the case $c_{1}=d_{1}=0$ and $c_{2}=d_{2}=100$, we obtain six fundamental rogue waves arraying a pentagon; among the six first-order rogue waves, one sits in the center and the rest are located on the vertices of the pentagon. If we set $c_{1}=d_{1}=c_{2}=d_{2}=100$, we observe also six fundamental rogue waves scattered in all directions,  arraying a regular triangle.

 The first and the secund-order rogue wave solutions illustrated in this paper resemble those obtained in Ref \cite{R20}. In this reference paper, the authors used a non-recursive DT. At our knowledge, the third-order rogue wave solutions for the Manakov system and a generalized formula for generating more than  third-order have not yet been constructed. Thus, we actually believe that the present work is worth underlying and paves ways for further investigations of the system above within the viewpoint of the existence of propagation of extreme wave excitations. The observation of these rogue waves profiles is expected to be possible in Manakov or near-Manakov systems such as birefringent fiber transmission links (\cite{R22},\cite{R23}-\cite{R25} ), crossing sea waves \cite{R17}, photorefractive media \cite{R27}, AlGaAs planar waveguide \cite{R26} and so on.

 Moreover, the results obtained here can be helpful to describe the oceanic rogue waves in  deep oceans and are expected to be useful for the technology of signal transmission through rogue wave. Further possible applications of our results can be found in the explanation of high intensity orthogonally polarized rogue light wave pulses in optical fibers. We  hope  our  result  will  be  realized  by  physical experiments in the future, which would be useful in the understanding of the generation mechanisms.

\section{ Appendix:}
 \textbf{Analytical expressions of coefficients of equation (\ref{Eq31})}

  $r_{1}^{[0]}=\frac {\sqrt {2}} {2\sqrt [4]{\beta}}\left( 1-i \right)  ( -2\,\sqrt {\beta}x
+3\,{\beta}^{3/2}\alpha\,t-4\,i\beta\,\alpha\,t-1 )\\
 {{\rm e}^{-2\,
\sqrt {\beta}x+2\,{\beta}^{3/2}\alpha\,t-4\,it\alpha\,\beta+1/2\,i
\beta\,t}}$,

$s_{1}^{[0]}=\frac {a\sqrt {2} }{{2\beta}^{3/4}}\left( 1-i \right)  ( -2\,\sqrt {\beta}x
+3\,{\beta}^{3/2}\alpha\,t-4\,i\beta\,\alpha\,t+1 ) \\
{{\rm e}^{-2\,
 \sqrt {\beta}x+2\,{\beta}^{3/2}\alpha\,t-4\,it\alpha\,\beta-1/2\,i
 \beta\,t}}$,

 $w_{1}^{[0]}=\frac {b\sqrt {2} }{{2\beta}^{3/4}}\left( 1-i \right)  ( -2\,\sqrt {\beta}x
+3\,{\beta}^{3/2}\alpha\,t-4\,i\beta\,\alpha\,t+1 )\\
 {{\rm e}^{-2\,
 \sqrt {\beta}x+2\,{\beta}^{3/2}\alpha\,t-4\,it\alpha\,\beta-1/2\,i
 \beta\,t}}$.

For simplicity, we have put  $\alpha=1$,($a=1$, $b=0$ for $s_{1}^{[j]}$),($a=0$, $b=1$ for $w_{1}^{[j]}$) (j=1,2)

$r_{1}^{[1]}=\left( -1/24+1/24\,i \right) \sqrt {2}{{\rm e}^{-2\,x+2\,t-7/2\,it}}
 ( -3+117\,{t}^{3}+180\,xt+15\,t-42\,x{t}^{2}+147\,{t}^{2}$\\
 $-18\,x-36\,{x}^{2}-36\,{x}^{2}t+24\,i{\it d_{1}}-240\,ixt-36\,it+504\,i{t}^{2}-144\,i{t}^{2}x+48\,i{x}^{2}t+44\,i{t}^{3}+24\,{\it c_{1}}+8\,{x}^{3}
 )$,

 $s_{1}^{[1]}= \left( 1/24-1/24\,i \right) \sqrt {2}\\
 {{\rm e}^{-2\,x+2\,t-9/2\,it}}
 ( -3-117\,{t}^{3}-252\,xt+129\,t+42\,x{t}^{2}-189\,{t}^{2}-30\,x+60\,{x}^{2}+36\,{x}^{2}t-24\,i{\it d_{1}}+336\,ixt-156\,it-648\,i{t}^{2}
+144\,i{t}^{2}x-24\,{\it c_{1}}-8\,{x}^{3}-48\,i{x}^{2}t-44\,i{t}^{3})$,

$w_{1}^{[1]}= \left( 1/24-1/24\,i \right) \sqrt {2}\\
{{\rm e}^{-2\,x+2\,t-9/2\,it}}
 ( -3-117\,{t}^{3}-252\,xt+129\,t+42\,x{t}^{2}-189\,{t}^{2}-30\,x+60\,{x}^{2}+36\,{x}^{2}t-24\,i{\it d_{1}}+336\,ixt-156\,it-648\,i{t}^{2}
+144\,i{t}^{2}x-24\,{\it c_{1}}-8\,{x}^{3}-48\,i{x}^{2}t-44\,i{t}^{3})$.

$r_{1}^{[2]}=\left( {\frac {1}{960}}-{\frac {1}{960}}\,i \right) \sqrt {2}\\
{{\rm e}^{-2\,x+2\,t-7/2\,it}} ( -45-210\,x-765\,t+11640\,{x}^{2}t-3720\,xt+21660\,x{t}^{2}+240\,{x}^{4}t+60840\,{t}^{3}x-4320\,t{x}^{3}-9240\,{t}^{2}{x}^{2}+5270\,{t}^{4}x
+560\,{t}^{2}{x}^{3}-4680\,{t}^{3}{x}^{2}-600\,{x}^{2}+3330\,{t}^{2}-1200\,{x}^{3}-102630\,{t}^{3}
-39525\,{t}^{4}+560\,{x}^{4}-237\,{t}^{5}-32\,{x}^{5}-3840\,ixtc_{{1}}+960\,ixd_{{1}}+22880\,ix{t}^{3}+1680\,i{t}^{2}d_{{1}}+1920\,i{t}^{2}{x}^{3}
+81120\,i{t}^{2}x+480\,ixt+5760\,i{x}^{3}t+5760\,itc_{{1}}+5760\,i{t}^{2}c_{{1}}+2880\,xc_{{1}}t+25200\,i{t}^{4}
+3116\,i{t}^{5}+540\,it+3840\,xd_{{1}}t-320\,i{x}^{4}t-31680\,i{t}^{2}{x}^{2}-14880\,i{x}^{2}t-4320\,itd_{{1}}-3360\,i{t}^{4}x-1760\,i{t}^{3}{x}^{2}-960\,i{x}^{2}d_{{1}}-5760\,td_{{1}}-4320\,tc_{{1}}-5760\,{t}^{2}d_{{1}}\\
+960\,xc_{{1}}+1680\,{t}^{2}c_{{1}}-960\,{x}^{2}c_{{1}}-41160\,i{t}^{3}-19440\,i{t}^{2}-240\,id_{{1}}-960\,id_{{2}}+2880\,ixtd_{{1}}-240\,c_{{1}}-960\,c_{{2}})$,

$s_{1}^{[2]}= \left( {\frac {1}{960}}-{\frac {1}{960}}\,i \right) \sqrt {2}\\
{{\rm e}^{-2\,x+2\,t-9/2\,it}} ( 45+270\,x+3555\,t+23160\,{x}^{2}t
-19800\,xt+38460\,x{t}^{2}+240\,{x}^{4}t+70200\,{t}^{3}x-5280\,t{x}^{3}-10920\,{t}^{2}{x}^{2}+5270\,{t}^{4}x+560\,{t}^{2}{x}^{3}-4680\,{t}^{3}{x}^{2}+1560\,{x}^{2}-12450\,{t}^{2}-3120\,{x}^{3}-158790\,{t}^{3}-44795\,{t}^{4}+720\,{x}^{4}-237\,{t}^{5}-32\,{x}^{5}
-30240\,i{x}^{2}t-7200\,itd_{{1}}-37440\,i{t}^{2}{x}^{2}+28560\,i{t}^{4}+2880\,ixd_{{1}}+20640\,ixt+9600\,itc_{{1}}+7040\,i{x}^{3}t+26400\,ix{t}^{3}+138720\,i{t}^{2}x-3840\,ixtc_{{1}}+1680\,i{t}^{2}d_{{1}}+1920\,i{t}^{2}{x}^{3}+5760\,i{t}^{2}c_{{1}}+2880\,xc_{{1}}t
+3116\,i{t}^{5}+3840\,xd_{{1}}t-320\,i{x}^{4}t-3360\,i{t}^{4}x-1760\,i{t}^{3}{x}^{2}-960\,i{x}^{2}d_{{1}}-62280\,i{t}^{3}
-1380\,it-80400\,i{t}^{2}-9600\,td_{{1}}-7200\,tc_{{1}}-5760\,{t}^{2}d_{{1}}+2880\,xc_{{1}}+1680\,{t}^{2}c_{{1}}-960\,{x}^{2}c_{{1}}-240\,id_{{1}}-960\,id_{{2}}+2880\,ixtd_{{1}}-240\,c_{{1}}-960\,c_{{2}})$,

$w_{1}^{[2]}= \left( {\frac {1}{960}}-{\frac {1}{960}}\,i \right) \sqrt {2}\\
{{\rm e}^{-2\,x+2\,t-9/2\,it}} ( 45+270\,x+3555\,t
+23160\,{x}^{2}t-19800\,xt+38460\,x{t}^{2}+240\,{x}^{4}t+70200\,{t}^{3}x-5280\,t{x}^{3}-10920\,{t}^{2}{x}^{2}
+5270\,{t}^{4}x+560\,{t}^{2}{x}^{3}-4680\,{t}^{3}{x}^{2}+1560\,{x}^{2}-12450\,{t}^{2}-3120\,{x}^{3}-158790\,{t}^{3}-44795\,{t}^{4}+720\,{x}^{4}-237\,{t}^{5}
-32\,{x}^{5}-30240\,i{x}^{2}t-7200\,itd_{{1}}-37440\,i{t}^{2}{x}^{2}+28560\,i{t}^{4}+2880\,ixd_{{1}}+20640\,ixt+9600\,itc_{{1}}+7040\,i{x}^{3}t+26400\,ix{t}^{3}+138720\,i{t}^{2}x-3840\,ixtc_{{1}}+1680\,i{t}^{2}d_{{1}}+1920\,i{t}^{2}{x}^{3}+5760\,i{t}^{2}c_{{1}}+2880\,xc_{{1}}t\\
+3116\,i{t}^{5}+3840\,xd_{{1}}t-320\,i{x}^{4}t-3360\,i{t}^{4}x-1760\,i{t}^{3}{x}^{2}-960\,i{x}^{2}d_{{1}}
-62280\,i{t}^{3}-1380\,it-80400\,i{t}^{2}-9600\,td_{{1}}-7200\,tc_{{1}}-5760\,{t}^{2}d_{{1}}+2880\,xc_{{1}}
+1680\,{t}^{2}c_{{1}}-960\,{x}^{2}c_{{1}}-240\,id_{{1}}-960\,id_{{2}}+2880\,ixtd_{{1}}-240\,c_{{1}}-960\,c_{{2}} )$.

\textbf{Analytical expressions for  } $r_{1}[1]$, $s_{1}[1]$, $w_{1}[1]$, $r_{1}[2]$, $s_{1}[2]$ and $w_{1}[2]$.

$r_{1}[1]=\frac { \left( 1/6-1/6\,i \right)\sqrt {2}}{
 \left( -8\,{x}^{2}+24\,xt-50\,{t}^{2}-2 \right)  \left( -2\,x+3\,t+4
\,it-1 \right) } \\
{{\rm e}^{-2\,x+2\,t-7/2\,it}}( -6+480\,it{x}^{4}+108\,t-24\,x-6100\,i{t}^{4}x-1952\,i{t}^{2}{x}^{3}-168\,i{t}^{2}{\it c_{1}}-576\,i{t}^{2}{\it d_{1}}-480\,i{x}^{2}t+384\,ix{\it d_{1}}\,t+24\,{\it d_{1}}-288\,ixt{\it c_{1}}+96\,i{x}^{2}{\it c_{1}}+1776\,ix{t}^{2}+384\,x{\it c_{1}}\,t +4464\,i{t}^{3}{x}^{2}-1032\,i{t}^{3}-64\,i{x}^{5}-12\,ix-24\,i{\it c1}+288\,xt{\it d_{1}}-3750\,{t}^{4}-48\,{x}^{2}+2900\,{t}^{3}+5000\,{t}^{5}-96\,{x}^{4}-300\,{t}^{2}-96\,{x}^{3}+3750\,i{t}^{5}+64\,i{x}^{3}+66\,it+128\,t{x}^{4}+168\,{t}^{2}{\it d_{1}}-4800\,{t}^{4}x-768\,{t}^{2}{x}^{3}-1224\,x{t}^{2}-576\,{t}^{2}{\it c_{1}}+2752\,{t}^{3}{x}^{2}+240\,{x}^{2}t-2064\,{x}^{2}{t}^{2}+3600\,x{t}^{3}-96\,{x}^{2}{\it d1}+576\,{x}^{3}t+144\,xt ) $,

$s_{1}[1]=\frac { \left( -1/6-1/6\,i \right) \sqrt {2}}{-8
\,{x}^{2}+24\,xt-50\,{t}^{2}-2}{{\rm e}^{-2\,x+2\,t-9/2\,it}} ( 12\,ix+42\,t-12\,x-24\,{\it c_{1}}
-350\,{t}^{4}-24\,{x}^{2}-366\,{t}^{3}+32\,{x}^{4}-486\,{t}^{2}+16\,{x}^{3}+204\,x{t}^{2}-168\,{x}^{2}t+432\,{x}^{2}{t}^{2}
-432\,x{t}^{3}-192\,{x}^{3}t+168\,xt+128\,i{x}^{3}t+96\,it{\it c_{1}}+48\,i{x}^{3}+72\,t{\it c_{1}}-96\,t{\it d1}-1200\,i{t}^{4}
-24\,i{x}^{2}-198\,i{t}^{2}-90\,it-362\,i{t}^{3}-24\,i{\it d_{1}}-576\,i{t}^{2}{x}^{2}-48\,ix{\it d_{1}}-120\,it{x}^{2}-24\,ixt
+72\,it{\it d_{1}}+228\,ix{t}^{2}+1376\,i{t}^{3}x-6\,i-48\,x{\it c_{1}} )$,

$w_{1}[1]=\frac { \left( -1/6-1/6\,i \right) \sqrt {2}}{-8
\,{x}^{2}+24\,xt-50\,{t}^{2}-2}{{\rm e}^{-2\,x+2\,t-9/2\,it}} ( 12\,ix+42\,t-12\,x-24\,{\it c_{1}}
-350\,{t}^{4}-24\,{x}^{2}-366\,{t}^{3}+32\,{x}^{4}-486\,{t}^{2}+16\,{x}^{3}+204\,x{t}^{2}-168\,{x}^{2}t+432\,{x}^{2}{t}^{2}
-432\,x{t}^{3}-192\,{x}^{3}t+168\,xt+128\,i{x}^{3}t+96\,it{\it c_{1}}+48\,i{x}^{3}+72\,t{\it c_{1}}-96\,t{\it d1}-1200\,i{t}^{4}
-24\,i{x}^{2}-198\,i{t}^{2}-90\,it-362\,i{t}^{3}-24\,i{\it d_{1}}-576\,i{t}^{2}{x}^{2}-48\,ix{\it d_{1}}-120\,it{x}^{2}-24\,ixt
+72\,it{\it d_{1}}+228\,ix{t}^{2}+1376\,i{t}^{3}x-6\,i-48\,x{\it c_{1}} )$,

$r_{1}[2]= \left( -1/30-1/30\,i \right)\\
 \sqrt {2}{{\rm e}^{ ( -{
\frac {2}{65}}+{\frac {7}{130}}\,i )  ( 16\,x+28\,ix-65\,t
 ) }} ( -540\,{\it d_{1}}+27648\,i{x}^{5}{\it d_{1}}\,t+540\,x-3510\,t
 -4860\,{x}^{2}t+14040\,xt+1890\,x{t}^{2}+9408\,{x}^{6}t-86016\,{x}^{5}{t}^{3}-12288\,{x}^{7}t+44544\,{x}^{6}{t}^{2}-71136\,{x}^{5}{t}
^{2}+36720\,{t}^{2}{{\it d_{1}}}^{2}-66960\,{t}^{2}{{\it c_{1}}}^{2}+8640\,{
x}^{2}{{\it c_{1}}}^{2}-350280\,{t}^{4}{x}^{3}+436068\,{t}^{5}{x}^{2}-
124800\,{t}^{7}x+60672\,{t}^{5}{x}^{3}-140448\,{t}^{6}{x}^{2}-1313928
\,{t}^{5}x-1694730\,{t}^{6}x+83520\,{x}^{3}{t}^{3}-10368\,{x}^{5}t-
36000\,{x}^{4}{t}^{2}+48960\,{t}^{4}{x}^{4}+35280\,{x}^{4}t-437760\,{t
}^{3}x-17280\,t{x}^{3}+218160\,{t}^{2}{x}^{2}-2115270\,{t}^{4}x-126000
\,{t}^{2}{x}^{3}+628200\,{t}^{3}{x}^{2}+2880\,{\it d_{1}}\,{x}^{4}-8640\,
{{\it d_{1}}}^{2}t+8640\,{x}^{2}{{\it d1}}^{2}+532440\,{x}^{2}{t}^{4}-
234720\,{t}^{4}{\it c_{1}}+111852\,{t}^{5}{\it d_{1}}-129564\,{t}^{5}{\it c_{1}
}-8064\,{x}^{5}{\it c_{1}}+432000\,{\it c_{1}}\,{t}^{3}-73440\,t{{\it c_{1}}}^{
2}+8640\,x{{\it c1}}^{2}-29160\,{t}^{2}{\it d_{1}}+244080\,{t}^{2}{\it c_{1}
}+8640\,{x}^{2}{\it c_{1}}-16740\,t{\it d_{1}}+15660\,t{\it c_{1}}-3240\,x{\it
c1}+29160\,{t}^{3}{\it d_{1}}+985140\,{t}^{4}{\it d_{1}}+276720\,{x}^{4}{t}^
{3}+4320\,{x}^{2}{\it d_{1}}-145152\,{x}^{6}{t}^{3}+277440\,{x}^{5}{t}^{4
}-6912\,{x}^{8}t+41472\,{x}^{7}{t}^{2}+168480\,{t}^{3}{{\it d_{1}}}^{2}-
121248\,{t}^{5}{x}^{4}-878176\,{t}^{6}{x}^{3}+11520\,{x}^{3}{{\it d_{1}}}
^{2}+2523600\,{t}^{7}{x}^{2}-3288750\,{t}^{8}x-123552\,{t}^{6}{\it d_{1}}
-1078536\,{t}^{6}{\it c1}-4608\,{x}^{6}{\it c1}+1080\,x{\it d1}+8640\,
{x}^{3}{\it d_{2}}+2160\,x{\it d_{2}}+1080\,{\it c_{1}}-1080\,{\it d_{2}}-1080\,{x
}^{2}-53190\,{t}^{2}+2520\,{x}^{3}-170325\,{t}^{3}-84480\,{t}^{3}{x}^{
3}{\it d_{1}}-718560\,{t}^{4}{x}^{2}{\it c_{1}}+311040\,{t}^{3}{x}^{3}{\it
c_{1}}+448704\,{t}^{5}x{\it d_{1}}+1315872\,{t}^{5}x{\it c_{1}}-9216\,{x}^{5}t{
\it d_{1}}-241920\,{t}^{4}{x}^{2}{\it d_{1}}+69120\,{x}^{4}{t}^{2}{\it d_{1}}+
41472\,{x}^{5}t{\it c_{1}}-155520\,{x}^{4}{t}^{2}{\it c_{1}}-51840\,t{{\it
d_{1}}}^{2}{x}^{2}-60480\,{t}^{2}{{\it d_{1}}}^{2}x+5760\,{x}^{3}t{\it d_{1}}-
25920\,x{\it c_{1}}\,t-21600\,x{\it d_{1}}\,t+108000\,{x}^{2}t{\it d_{1}}-69120
\,{x}^{2}t{\it c_{1}}+262440\,{t}^{4}x{\it c_{1}}+8640\,{x}^{4}t{\it d_{1}}+
48960\,{x}^{4}t{\it c_{1}}+538560\,{\it c_{1}}\,{t}^{3}x+11520\,{\it c_{1}}\,t{
x}^{3}-190080\,{\it c_{1}}\,{t}^{2}{x}^{2}+8640\,t{{\it c_{1}}}^{2}x-60480\,
x{{\it d_{1}}}^{2}t-943200\,x{t}^{3}{\it d_{1}}-438480\,x{t}^{2}{\it d_{1}}+
73440\,{x}^{2}{t}^{2}{\it d_{1}}-358920\,{t}^{4}x{\it d_{1}}-182880\,{t}^{3}
{x}^{2}{\it c_{1}}+31680\,{t}^{2}{x}^{3}{\it d_{1}}-66240\,{t}^{2}{x}^{3}{
\it c_{1}}+27360\,{t}^{3}{x}^{2}{\it d_{1}}-644130\,{t}^{4}-1440\,{x}^{4}+
1790073\,{t}^{5}-4320\,{x}^{5}+2688\,{x}^{6}-26250\,{t}^{8}+8682\,{t}^
{6}+2125893\,{t}^{7}+1536\,{x}^{8}-384\,{x}^{7}-2160\,{{\it d_{1}}}^{2}-
2160\,{{\it c_{1}}}^{2}+1828125\,{t}^{9}+512\,{x}^{9}-35640\,{t}^{2}{\it
d2}-4320\,{x}^{2}{\it c_{2}}+2880\,{x}^{3}{\it c_{2}}-8640\,{x}^{3}{\it d_{1}}-
5760\,{x}^{5}{\it d_{1}}-16200\,t{\it d_{2}}+7560\,t{\it c_{2}}-2160\,x{\it c_{2}}
+5760\,{x}^{4}{\it c_{2}}-87480\,{t}^{2}{\it c_{2}}-216000\,{t}^{4}{\it d_{2}}-
65880\,{t}^{3}{\it c_{2}}-63000\,{t}^{4}{\it c_{2}}-65160\,{t}^{3}{\it d_{2}}-
69120\,{x}^{2}t{\it c_{1}}\,{\it d_{1}}+207360\,{t}^{2}{\it c_{1}}\,x{\it d_{1}}+
51840\,xt{\it c_{1}}\,{\it d_{1}}-63360\,{t}^{3}{\it d_{1}}\,{\it c_{1}}-17280\,{x
}^{2}{\it c_{1}}\,{\it d_{1}}+8640\,x{\it c_{1}}\,{\it d_{1}}+12960\,t{\it c_{1}}\,{
\it c_{2}}+247680\,{t}^{3}x{\it d_{2}}+30240\,{t}^{2}{\it c_{1}}\,{\it d_{1}}-2160
\,ix{\it c_{2}}+23040\,{x}^{3}t{\it d_{2}}+36720\,x{t}^{2}{\it c_{2}}-4320\,xt{
\it d_{2}}-21600\,{x}^{2}t{\it d_{2}}-34560\,{x}^{3}t{\it c_{2}}+30240\,xt{\it
c2}-30240\,{x}^{2}t{\it c_{2}}-103680\,{t}^{2}{x}^{2}{\it d_{2}}+77760\,{t}^
{2}{x}^{2}{\it c_{2}}-77760\,{t}^{3}x{\it c_{2}}+12960\,t{\it d_{1}}\,{\it d_{2}}+
38880\,{\it c_{1}}\,t{\it d_{1}}+41040\,{t}^{2}x{\it d_{2}}+760320\,i{x}^{3}{t}
^{3}{\it d_{1}}+56160\,ixt{\it c_{1}}+77760\,i{t}^{2}{x}^{2}{\it d_{2}}+8640\,i
x{\it d_{1}}\,{\it c_{2}}+4320\,ixt{\it c_{2}}+60480\,ixt{{\it d1}}^{2}+21600\,
i{x}^{2}t{\it c_{2}}+64800\,ix{t}^{2}{\it d_{1}}+8640\,ix{{\it c_{1}}}^{2}t+
4320\,i{{\it d_{1}}}^{2}x+4320\,i{x}^{2}{\it c_{2}}+8640\,i{\it d_{1}}\,{{\it
c_{1}}}^{2}+244800\,i{\it c_{1}}\,{t}^{2}{x}^{3}+79920\,i{t}^{3}{\it d_{1}}+
4320\,i{\it d_{1}}\,{\it c_{2}}+228960\,i{t}^{4}{\it d_{1}}+23760\,it{{\it c_{1}}}
^{2}+62640\,i{{\it d_{1}}}^{2}t+35640\,i{t}^{2}{\it c_{2}}+5760\,i{x}^{4}{
\it d_{2}}+216000\,i{t}^{4}{\it c_{2}}+63360\,i{t}^{3}{{\it c_{1}}}^{2}+286200
\,i{t}^{2}{\it d_{1}}+4320\,i{x}^{2}{\it d_{1}}+226304\,i{x}^{6}{t}^{3}+
1582920\,i{t}^{4}{x}^{3}+32940\,i{x}^{2}t+315000\,i{t}^{3}{x}^{2}+
2292864\,i{t}^{5}{x}^{4}+112608\,i{x}^{5}{t}^{2}+3072\,i{x}^{8}t+
4795200\,i{t}^{7}{x}^{2}+1102890\,i{t}^{6}x+19440\,i{x}^{4}t+65160\,i{
t}^{3}{\it c_{2}}+2036400\,i{t}^{7}x+89280\,i{t}^{3}x+15360\,i{x}^{7}t+
832104\,i{t}^{5}x+1802304\,i{t}^{5}{x}^{3}+563400\,i{x}^{2}{t}^{4}+
2880\,i{x}^{3}{\it d_{2}}+6480\,i{t}^{2}{x}^{2}+16200\,it{\it c_{2}}+7560\,i
t{\it d_{2}}+8640\,i{\it c_{1}}\,{x}^{3}+18720\,i{x}^{4}{t}^{2}+5760\,i{x}^{
5}t+398592\,i{x}^{5}{t}^{3}+1080\,ixt-23040\,i{x}^{3}t{\it c_{2}}-315360
\,i{\it c_{1}}\,{t}^{2}{x}^{2}-195840\,i{x}^{4}{t}^{2}{\it d_{1}}-168480\,i{
t}^{3}{\it d_{1}}\,{\it c_{1}}-207360\,i{t}^{2}{{\it c_{1}}}^{2}x-8640\,i{\it
c1}\,x{\it d_{2}}-241920\,i{x}^{2}{t}^{4}{\it c_{1}}-9216\,i{x}^{5}{\it c_{1}}
\,t-34560\,it{x}^{3}{\it d_{2}}-103680\,i{\it c_{1}}\,{t}^{2}{\it d_{1}}-8640\,
i{x}^{2}t{\it d_{1}}-41040\,i{t}^{2}x{\it c_{2}}-54720\,it{x}^{4}{\it c_{1}}-
124920\,i{t}^{4}{\it c_{1}}\,x-38880\,it{\it c_{1}}\,{x}^{2}-496440\,ix{t}^{
4}{\it d_{1}}-280800\,i{\it c_{1}}\,{t}^{3}{x}^{2}-146880\,i{t}^{2}{x}^{3}{
\it d_{1}}-30240\,i{x}^{2}t{\it d_{2}}+8640\,i{x}^{2}{{\it c_{1}}}^{2}+5760\,i{
x}^{5}{\it c_{1}}+149400\,i{\it c_{1}}\,{t}^{3}+336852\,i{t}^{5}{\it c_{1}}+
8100\,it{\it c_{1}}-247680\,i{t}^{3}x{\it c_{2}}-11520\,i{\it c_{1}}\,{x}^{3}{
\it d_{1}}-64800\,it{\it c_{1}}\,{\it d_{1}}-1477440\,i{x}^{2}{t}^{4}{\it d_{1}}-
84480\,i{x}^{3}{t}^{3}{\it c_{1}}-17280\,it{\it d_{1}}\,{\it d_{2}}-17280\,it{
\it c1}\,{\it c_{2}}-155520\,i{x}^{2}{t}^{2}{\it d_{1}}-12960\,ix{\it d_{1}}\,t
-12960\,it{\it d_{1}}\,{\it c_{2}}-77760\,i{t}^{3}x{\it d_{2}}-65880\,i{t}^{3}{
\it d_{2}}-66960\,i{t}^{2}{{\it c_{1}}}^{2}-8640\,i{x}^{3}{\it c_{2}}-95436\,i{
t}^{5}{\it d_{1}}-4320\,ix{{\it c_{1}}}^{2}-128250\,ix{t}^{2}-1392210\,i{t}^
{4}x-36864\,i{x}^{7}{t}^{2}-875520\,i{x}^{5}{t}^{4}-4108032\,i{t}^{6}{
x}^{3}-585840\,i{x}^{4}{t}^{3}-14016\,i{x}^{6}t-1901124\,i{t}^{5}{x}^{
2}-43920\,i{t}^{2}{x}^{3}-3240000\,i{t}^{8}x-3072\,i{x}^{6}{\it d_{1}}-
4320\,i{x}^{2}{\it c_{1}}-4320\,i{x}^{2}{\it d_{2}}-63000\,i{t}^{4}{\it d_{2}}-
160920\,i{t}^{2}{\it c_{1}}-2880\,i{x}^{4}{\it c_{1}}-774900\,i{\it c_{1}}\,{t}
^{4}-87480\,i{t}^{2}{\it d_{2}}-796464\,i{t}^{6}{\it d_{1}}-123552\,i{t}^{6}
{\it c_{1}}-36720\,i{t}^{2}{{\it d_{1}}}^{2}-4320\,i{\it c_{1}}\,{\it d_{2}}-3456
\,i{x}^{5}{\it d_{1}}-2160\,ix{\it d_{2}}-8640\,i{x}^{2}{{\it d_{1}}}^{2}-1080
\,i{\it c_{1}}\,x-1620\,it{\it d_{1}}-5760\,i{x}^{3}{\it d_{1}}-2251536\,i{t}^{
6}{x}^{2}-17280\,i{x}^{3}t-100608\,i{x}^{6}{t}^{2}-1013280\,i{x}^{4}{t
}^{4}-348480\,i{x}^{3}{t}^{3}-3240\,ix{\it d_{1}}+14400\,it{x}^{4}{\it d_{1}
}+17280\,i{\it c_{1}}\,{x}^{3}t+12960\,i{\it c_{1}}\,t{\it d_{2}}+448704\,i{t}^
{5}x{\it c_{1}}+69120\,i{x}^{2}{{\it c_{1}}}^{2}t+1080\,i{\it c_{2}}+8640\,i{{
\it d_{1}}}^{3}+69120\,i{x}^{4}{t}^{2}{\it c_{1}}+384\,i{x}^{7}+540\,i{\it
c_{1}}+1080\,i{\it d_{1}}+30240\,ixt{\it d_{2}}+69120\,it{\it c_{1}}\,x{\it d_{1}}+
51840\,i{x}^{2}{\it d_{1}}\,t{\it c_{1}}-135\,i+60480\,i{t}^{2}{\it d_{1}}\,x{
\it c_{1}}+92880\,i{\it c_{1}}\,{t}^{2}x+8640\,ix{\it c_{1}}\,{\it d_{1}}-4320\,{x
}^{2}{\it d_{2}}-4320\,{\it c_{1}}\,{\it c_{2}}-4320\,{\it d_{1}}\,{\it d_{2}}-768\,i
{x}^{8}-3960\,i{x}^{3}-270\,ix-3744\,i{x}^{5}-1094349\,i{t}^{7}-
1261875\,i{t}^{8}-327870\,i{t}^{4}-2128626\,i{t}^{6}-1152\,i{x}^{6}-
33750\,i{t}^{2}-1080\,i{x}^{2}+8640\,{\it c_{1}}\,{{\it d_{1}}}^{2}+2880\,ix
{t}^{3}{\it d_{1}}+36720\,ix{t}^{2}{\it d_{2}}+687500\,i{t}^{9}+457065\,i{t}
^{3}+1135251\,i{t}^{5}+1485\,it+1440\,i{x}^{4}-8640\,x{\it d_{1}}\,{\it
d_{2}}-8640\,x{\it c_{1}}\,{\it c_{2}}-17280\,t{\it d_{1}}\,{\it c_{2}}+17280\,{\it
c_{1}}\,t{\it d_{2}}+1236960\,i{\it c_{1}}\,{t}^{3}x+103680\,i{t}^{2}{x}^{2}{
\it c_{2}}+8640\,{{\it c_{1}}}^{3}+491040\,i{t}^{3}{x}^{2}{\it d_{1}}+1384128\,
ix{t}^{5}{\it d_{1}}+34560\,it{\it d_{1}}\,{x}^{3} )/(9-612\,xt-192\,
{x}^{3}{\it c_{1}}-22500\,{t}^{5}x-8928\,{x}^{3}{t}^{3}-576\,{x}^{5}t+
2928\,{x}^{4}{t}^{2}+792\,{t}^{3}x+96\,t{x}^{3}-2232\,{t}^{2}{x}^{2}+
18300\,{x}^{2}{t}^{4}-2808\,{\it c_{1}}\,{t}^{3}+864\,t{\it d_{1}}-792\,t{
\it c_{1}}+144\,x{\it c_{1}}-1056\,{t}^{3}{\it d_{1}}+108\,{x}^{2}+2835\,{t}^{2
}-1152\,{x}^{2}t{\it d_{1}}+864\,{x}^{2}t{\it c_{1}}+1008\,{\it c_{1}}\,x{t}^{2
}+3456\,x{t}^{2}{\it d_{1}}+10179\,{t}^{4}+48\,{x}^{4}+64\,{x}^{6}+15625
\,{t}^{6}+144\,{{\it d_{1}}}^{2}+144\,{{\it c_{1}}}^{2})$,

$s_{1}[2]= \left( -1/30-1/30\,i \right) \\
\sqrt {2}{{\rm e}^{ \left( -{
\frac {2}{97}}+{\frac {9}{194}}\,i \right)  ( 16\,x+36\,ix-97\,t
 ) }}( -540\,{\it d_{1}}+95436\,i{t}^{5}{\it d_{1}}+8640\,i{x}^{
3}{\it c_{2}}+66960\,i{t}^{2}{{\it c_{1}}}^{2}+65880\,i{t}^{3}{\it d_{2}}+3240
\,ix{\it d_{1}}+348480\,i{x}^{3}{t}^{3}+1013280\,i{x}^{4}{t}^{4}+100608\,
i{x}^{6}{t}^{2}+17280\,i{x}^{3}t+2251536\,i{t}^{6}{x}^{2}+5760\,i{x}^{
3}{\it d_{1}}+1620\,it{\it d_{1}}+1080\,i{\it c_{1}}\,x+2160\,ix{\it d_{2}}+3456\,
i{x}^{5}{\it d_{1}}+4320\,i{\it c_{1}}\,{\it d_{2}}+36720\,i{t}^{2}{{\it d_{1}}}^{
2}-491040\,i{t}^{3}{x}^{2}{\it d_{1}}-244800\,i{\it c_{1}}\,{t}^{2}{x}^{3}-
92880\,i{\it c_{1}}\,{t}^{2}x-8640\,ix{{\it c_{1}}}^{2}t+27648\,i{x}^{5}{
\it d_{1}}\,t+103680\,i{\it c_{1}}\,{t}^{2}{\it d_{1}}+41040\,i{t}^{2}x{\it c_{2}}
+540\,x-3510\,t+8640\,i{x}^{2}t{\it d_{1}}+496440\,ix{t}^{4}{\it d_{1}}+
38880\,it{\it c_{1}}\,{x}^{2}+124920\,i{t}^{4}{\it c_{1}}\,x+54720\,it{x}^{4
}{\it c_{1}}+30240\,i{x}^{2}t{\it d_{2}}+146880\,i{t}^{2}{x}^{3}{\it d_{1}}+
280800\,i{\it c_{1}}\,{t}^{3}{x}^{2}-4860\,{x}^{2}t-14040\,xt+1890\,x{t}^
{2}+9408\,{x}^{6}t+86016\,{x}^{5}{t}^{3}+12288\,{x}^{7}t-44544\,{x}^{6
}{t}^{2}-71136\,{x}^{5}{t}^{2}-36720\,{t}^{2}{{\it d_{1}}}^{2}+66960\,{t}
^{2}{{\it c_{1}}}^{2}-8640\,{x}^{2}{{\it c_{1}}}^{2}-350280\,{t}^{4}{x}^{3}+
436068\,{t}^{5}{x}^{2}+124800\,{t}^{7}x-60672\,{t}^{5}{x}^{3}+140448\,
{t}^{6}{x}^{2}+1313928\,{t}^{5}x-1694730\,{t}^{6}x-83520\,{x}^{3}{t}^{
3}+10368\,{x}^{5}t+36000\,{x}^{4}{t}^{2}-48960\,{t}^{4}{x}^{4}+35280\,
{x}^{4}t+437760\,{t}^{3}x+17280\,t{x}^{3}-218160\,{t}^{2}{x}^{2}-
2115270\,{t}^{4}x-126000\,{t}^{2}{x}^{3}+628200\,{t}^{3}{x}^{2}+2880\,
{\it d_{1}}\,{x}^{4}-8640\,{{\it d1}}^{2}t-8640\,{x}^{2}{{\it d_{1}}}^{2}-
532440\,{x}^{2}{t}^{4}-234720\,{t}^{4}{\it c_{1}}-111852\,{t}^{5}{\it d_{1}}
+129564\,{t}^{5}{\it c_{1}}+8064\,{x}^{5}{\it c_{1}}-432000\,{\it c_{1}}\,{t}^{
3}-73440\,t{{\it c_{1}}}^{2}+8640\,x{{\it c_{1}}}^{2}-29160\,{t}^{2}{\it d_{1}}
+244080\,{t}^{2}{\it c_{1}}+8640\,{x}^{2}{\it c_{1}}+16740\,t{\it d_{1}}-15660
\,t{\it c_{1}}+3240\,x{\it c_{1}}-29160\,{t}^{3}{\it d_{1}}+985140\,{t}^{4}{
\it d_{1}}+276720\,{x}^{4}{t}^{3}+4320\,{x}^{2}{\it d_{1}}-145152\,{x}^{6}{t
}^{3}+277440\,{x}^{5}{t}^{4}-6912\,{x}^{8}t+41472\,{x}^{7}{t}^{2}+
168480\,{t}^{3}{{\it d_{1}}}^{2}-121248\,{t}^{5}{x}^{4}-878176\,{t}^{6}{x
}^{3}+11520\,{x}^{3}{{\it d_{1}}}^{2}+2523600\,{t}^{7}{x}^{2}-3288750\,{t
}^{8}x-123552\,{t}^{6}{\it d_{1}}-1078536\,{t}^{6}{\it c_{1}}-4608\,{x}^{6}{
\it c_{1}}-1080\,x{\it d_{1}}-8640\,{x}^{3}{\it d_{2}}-2160\,x{\it d_{2}}+1080\,{
\it c_{1}}-1080\,{\it d_{2}}+1080\,{x}^{2}+53190\,{t}^{2}+2520\,{x}^{3}-
170325\,{t}^{3}-84480\,{t}^{3}{x}^{3}{\it d_{1}}-718560\,{t}^{4}{x}^{2}{
\it c_{1}}+311040\,{t}^{3}{x}^{3}{\it c_{1}}+448704\,{t}^{5}x{\it d1}+
1315872\,{t}^{5}x{\it c_{1}}-9216\,{x}^{5}t{\it d_{1}}-241920\,{t}^{4}{x}^{2
}{\it d_{1}}+69120\,{x}^{4}{t}^{2}{\it d_{1}}+41472\,{x}^{5}t{\it c_{1}}-155520
\,{x}^{4}{t}^{2}{\it c_{1}}-51840\,t{{\it d_{1}}}^{2}{x}^{2}-60480\,{t}^{2}{
{\it d_{1}}}^{2}x+5760\,{x}^{3}t{\it d_{1}}-25920\,x{\it c_{1}}\,t-21600\,x{
\it d_{1}}\,t-108000\,{x}^{2}t{\it d_{1}}+69120\,{x}^{2}t{\it c_{1}}-262440\,{t
}^{4}x{\it c_{1}}-8640\,{x}^{4}t{\it d_{1}}-48960\,{x}^{4}t{\it c_{1}}+538560\,
{\it c_{1}}\,{t}^{3}x+11520\,{\it c_{1}}\,t{x}^{3}-190080\,{\it c_{1}}\,{t}^{2}
{x}^{2}-8640\,t{{\it c1}}^{2}x+60480\,x{{\it d1}}^{2}t-943200\,x{t}^{3
}{\it d_{1}}+438480\,x{t}^{2}{\it d_{1}}+73440\,{x}^{2}{t}^{2}{\it d_{1}}+
358920\,{t}^{4}x{\it d_{1}}+182880\,{t}^{3}{x}^{2}{\it c_{1}}-31680\,{t}^{2}
{x}^{3}{\it d_{1}}+66240\,{t}^{2}{x}^{3}{\it c_{1}}-27360\,{t}^{3}{x}^{2}{
\it d_{1}}+644130\,{t}^{4}+1440\,{x}^{4}+1790073\,{t}^{5}-4320\,{x}^{5}-
2688\,{x}^{6}+26250\,{t}^{8}-8682\,{t}^{6}+2125893\,{t}^{7}-1536\,{x}^
{8}-384\,{x}^{7}+2160\,{{\it d_{1}}}^{2}+2160\,{{\it c_{1}}}^{2}+1828125\,{t
}^{9}+512\,{x}^{9}-35640\,{t}^{2}{\it d_{2}}-4320\,{x}^{2}{\it c_{2}}-2880\,
{x}^{3}{\it c_{2}}+8640\,{x}^{3}{\it d_{1}}+5760\,{x}^{5}{\it d_{1}}+16200\,t{
\it d_{2}}-7560\,t{\it c_{2}}+2160\,x{\it c_{2}}+5760\,{x}^{4}{\it c_{2}}-87480\,{
t}^{2}{\it c_{2}}-216000\,{t}^{4}{\it d_{2}}+65880\,{t}^{3}{\it c_{2}}-63000\,{
t}^{4}{\it c_{2}}+65160\,{t}^{3}{\it d_{2}}-69120\,{x}^{2}t{\it c_{1}}\,{\it d_{1}
}+207360\,{t}^{2}{\it c_{1}}\,x{\it d_{1}}-60480\,ixt{{\it d_{1}}}^{2}-51840\,x
t{\it c_{1}}\,{\it d_{1}}-63360\,{t}^{3}{\it d_{1}}\,{\it c_{1}}+17280\,{x}^{2}{
\it c_{1}}\,{\it d_{1}}+8640\,x{\it c_{1}}\,{\it d_{1}}+12960\,t{\it c_{1}}\,{\it c_{2}}
+247680\,{t}^{3}x{\it d_{2}}-30240\,{t}^{2}{\it c_{1}}\,{\it d_{1}}+8640\,i{x}^
{2}{{\it d_{1}}}^{2}-8640\,i{x}^{2}{{\it c_{1}}}^{2}+23040\,{x}^{3}t{\it d_{2}}
-36720\,x{t}^{2}{\it c_{2}}-4320\,xt{\it d_{2}}+21600\,{x}^{2}t{\it d_{2}}-
34560\,{x}^{3}t{\it c_{2}}+30240\,xt{\it c_{2}}+30240\,{x}^{2}t{\it c_{2}}-
103680\,{t}^{2}{x}^{2}{\it d_{2}}+77760\,{t}^{2}{x}^{2}{\it c_{2}}-77760\,{t
}^{3}x{\it c_{2}}+12960\,t{\it d_{1}}\,{\it d_{2}}+38880\,{\it c_{1}}\,t{\it d_{1}}-
41040\,{t}^{2}x{\it d_{2}}+760320\,i{x}^{3}{t}^{3}{\it d_{1}}-21600\,i{x}^{2
}t{\it c_{2}}-14400\,it{x}^{4}{\it d_{1}}-64800\,ix{t}^{2}{\it d_{1}}-36720\,ix
{t}^{2}{\it d_{2}}+56160\,ixt{\it c_{1}}+77760\,i{t}^{2}{x}^{2}{\it d_{2}}+8640
\,ix{\it d_{1}}\,{\it c_{2}}+4320\,ixt{\it c_{2}}-1440\,i{x}^{4}+1080\,i{x}^{2}
+33750\,i{t}^{2}+1152\,i{x}^{6}+2128626\,i{t}^{6}+327870\,i{t}^{4}+
1261875\,i{t}^{8}+768\,i{x}^{8}-69120\,it{\it c_{1}}\,x{\it d_{1}}+4320\,i{{
\it d_{1}}}^{2}x+4320\,i{x}^{2}{\it c_{2}}+8640\,i{\it d_{1}}\,{{\it c_{1}}}^{2}+
228960\,i{t}^{4}{\it d_{1}}+23760\,it{{\it c_{1}}}^{2}+62640\,i{{\it d_{1}}}^{2
}t+35640\,i{t}^{2}{\it c_{2}}+5760\,i{x}^{4}{\it d_{2}}+216000\,i{t}^{4}{
\it c_{2}}+63360\,i{t}^{3}{{\it c_{1}}}^{2}+286200\,i{t}^{2}{\it d_{1}}+4320\,i
{x}^{2}{\it d_{1}}+226304\,i{x}^{6}{t}^{3}+1582920\,i{t}^{4}{x}^{3}+32940
\,i{x}^{2}t+315000\,i{t}^{3}{x}^{2}+2292864\,i{t}^{5}{x}^{4}+112608\,i
{x}^{5}{t}^{2}+3072\,i{x}^{8}t+4795200\,i{t}^{7}{x}^{2}+1102890\,i{t}^
{6}x+19440\,i{x}^{4}t-23040\,i{x}^{3}t{\it c_{2}}-315360\,i{\it c_{1}}\,{t}^
{2}{x}^{2}-195840\,i{x}^{4}{t}^{2}{\it d_{1}}-168480\,i{t}^{3}{\it d_{1}}\,{
\it c_{1}}-207360\,i{t}^{2}{{\it c_{1}}}^{2}x-8640\,i{\it c_{1}}\,x{\it d_{2}}-
241920\,i{x}^{2}{t}^{4}{\it c_{1}}-9216\,i{x}^{5}{\it c_{1}}\,t-34560\,it{x}
^{3}{\it d_{2}}-247680\,i{t}^{3}x{\it c_{2}}-11520\,i{\it c_{1}}\,{x}^{3}{\it
d_{1}}-64800\,it{\it c_{1}}\,{\it d_{1}}-1477440\,i{x}^{2}{t}^{4}{\it d_{1}}-84480
\,i{x}^{3}{t}^{3}{\it c_{1}}-17280\,it{\it d_{1}}\,{\it d_{2}}-17280\,it{\it c_{1}
}\,{\it c_{2}}-155520\,i{x}^{2}{t}^{2}{\it d_{1}}-12960\,ix{\it d_{1}}\,t-12960
\,it{\it d_{1}}\,{\it c_{2}}-77760\,i{t}^{3}x{\it d_{2}}-4320\,ix{{\it c_{1}}}^{2}
-128250\,ix{t}^{2}-1392210\,i{t}^{4}x-36864\,i{x}^{7}{t}^{2}-875520\,i
{x}^{5}{t}^{4}-4108032\,i{t}^{6}{x}^{3}-585840\,i{x}^{4}{t}^{3}-14016
\,i{x}^{6}t-1901124\,i{t}^{5}{x}^{2}-43920\,i{t}^{2}{x}^{3}-3240000\,i
{t}^{8}x-3072\,i{x}^{6}{\it d_{1}}-4320\,i{x}^{2}{\it c_{1}}-4320\,i{x}^{2}{
\it d_{2}}-63000\,i{t}^{4}{\it d_{2}}-160920\,i{t}^{2}{\it c_{1}}-2880\,i{x}^{4
}{\it c_{1}}-774900\,i{\it c_{1}}\,{t}^{4}-87480\,i{t}^{2}{\it d_{2}}-796464\,i
{t}^{6}{\it d_{1}}-123552\,i{t}^{6}{\it c_{1}}+17280\,i{\it c_{1}}\,{x}^{3}t+
12960\,i{\it c_{1}}\,t{\it d_{2}}+448704\,i{t}^{5}x{\it c_{1}}+69120\,i{x}^{2}{
{\it c_{1}}}^{2}t+1080\,i{\it c_{2}}+8640\,i{{\it d_{1}}}^{3}+69120\,i{x}^{4}{t
}^{2}{\it c_{1}}+384\,i{x}^{7}+540\,i{\it c_{1}}+1080\,i{\it d_{1}}+30240\,ixt{
\it d_{2}}+51840\,i{x}^{2}{\it d_{1}}\,t{\it c_{1}}+60480\,i{t}^{2}{\it d_{1}}\,x{
\it c_{1}}+8640\,ix{\it c_{1}}\,{\it d_{1}}-4320\,{x}^{2}{\it d_{2}}+4320\,{\it c_{1}
}\,{\it c_{2}}+4320\,{\it d_{1}}\,{\it d_{2}}-3960\,i{x}^{3}-270\,ix-3744\,i{x}
^{5}-1094349\,i{t}^{7}+8640\,{\it c_{1}}\,{{\it d_{1}}}^{2}+2160\,ix{\it c_{2}}
+2880\,ix{t}^{3}{\it d_{1}}+687500\,i{t}^{9}+457065\,i{t}^{3}+1135251\,i{
t}^{5}+1485\,it+135\,i-8640\,x{\it d_{1}}\,{\it d_{2}}-8640\,x{\it c_{1}}\,{
\it c_{2}}-17280\,t{\it d_{1}}\,{\it c_{2}}+17280\,{\it c_{1}}\,t{\it d_{2}}+1236960
\,i{\it c_{1}}\,{t}^{3}x+103680\,i{t}^{2}{x}^{2}{\it c_{2}}-4320\,i{\it d_{1}}
\,{\it c_{2}}-6480\,i{t}^{2}{x}^{2}-563400\,i{x}^{2}{t}^{4}-1802304\,i{t}
^{5}{x}^{3}-832104\,i{t}^{5}x-15360\,i{x}^{7}t-89280\,i{t}^{3}x-
2036400\,i{t}^{7}x-1080\,ixt-398592\,i{x}^{5}{t}^{3}-5760\,i{x}^{5}t-
18720\,i{x}^{4}{t}^{2}-336852\,i{t}^{5}{\it c_{1}}-8100\,it{\it c_{1}}-65160
\,i{t}^{3}{\it c_{2}}-2880\,i{x}^{3}{\it d_{2}}-7560\,it{\it d_{2}}-16200\,it{
\it c_{2}}-5760\,i{x}^{5}{\it c_{1}}-8640\,i{\it c_{1}}\,{x}^{3}-149400\,i{\it
c_{1}}\,{t}^{3}-79920\,i{t}^{3}{\it d_{1}}+8640\,{{\it c_{1}}}^{3}+1384128\,ix{
t}^{5}{\it d_{1}}+34560\,it{\it d_{1}}\,{x}^{3} )/(9-612\,xt-192\,{x}
^{3}{\it c_{1}}-22500\,{t}^{5}x-8928\,{x}^{3}{t}^{3}-576\,{x}^{5}t+2928\,
{x}^{4}{t}^{2}+792\,{t}^{3}x+96\,t{x}^{3}-2232\,{t}^{2}{x}^{2}+18300\,
{x}^{2}{t}^{4}-2808\,{\it c_{1}}\,{t}^{3}+864\,t{\it d_{1}}-792\,t{\it c_{1}}+
144\,x{\it c_{1}}-1056\,{t}^{3}{\it d_{1}}+108\,{x}^{2}+2835\,{t}^{2}-1152\,
{x}^{2}t{\it d_{1}}+864\,{x}^{2}t{\it c_{1}}+1008\,{\it c_{1}}\,x{t}^{2}+3456\,
x{t}^{2}{\it d_{1}}+10179\,{t}^{4}+48\,{x}^{4}+64\,{x}^{6}+15625\,{t}^{6}
+144\,{{\it d_{1}}}^{2}+144\,{{\it c_{1}}}^{2})$,

$w_{1}[2]= \left( -1/30-1/30\,i \right) \\
\sqrt {2}{{\rm e}^{ \left( -{
\frac {2}{97}}+{\frac {9}{194}}\,i \right)  ( 16\,x+36\,ix-97\,t
 ) }}( -540\,{\it d_{1}}+95436\,i{t}^{5}{\it d_{1}}+8640\,i{x}^{
3}{\it c_{2}}+66960\,i{t}^{2}{{\it c_{1}}}^{2}+65880\,i{t}^{3}{\it d_{2}}+3240
\,ix{\it d_{1}}+348480\,i{x}^{3}{t}^{3}+1013280\,i{x}^{4}{t}^{4}+100608\,
i{x}^{6}{t}^{2}+17280\,i{x}^{3}t+2251536\,i{t}^{6}{x}^{2}+5760\,i{x}^{
3}{\it d_{1}}+1620\,it{\it d_{1}}+1080\,i{\it c_{1}}\,x+2160\,ix{\it d_{2}}+3456\,
i{x}^{5}{\it d_{1}}+4320\,i{\it c_{1}}\,{\it d_{2}}+36720\,i{t}^{2}{{\it d_{1}}}^{
2}-491040\,i{t}^{3}{x}^{2}{\it d_{1}}-244800\,i{\it c_{1}}\,{t}^{2}{x}^{3}-
92880\,i{\it c_{1}}\,{t}^{2}x-8640\,ix{{\it c_{1}}}^{2}t+27648\,i{x}^{5}{
\it d_{1}}\,t+103680\,i{\it c_{1}}\,{t}^{2}{\it d_{1}}+41040\,i{t}^{2}x{\it c_{2}}
+540\,x-3510\,t+8640\,i{x}^{2}t{\it d_{1}}+496440\,ix{t}^{4}{\it d_{1}}+
38880\,it{\it c_{1}}\,{x}^{2}+124920\,i{t}^{4}{\it c_{1}}\,x+54720\,it{x}^{4
}{\it c_{1}}+30240\,i{x}^{2}t{\it d_{2}}+146880\,i{t}^{2}{x}^{3}{\it d_{1}}+
280800\,i{\it c_{1}}\,{t}^{3}{x}^{2}-4860\,{x}^{2}t-14040\,xt+1890\,x{t}^
{2}+9408\,{x}^{6}t+86016\,{x}^{5}{t}^{3}+12288\,{x}^{7}t-44544\,{x}^{6
}{t}^{2}-71136\,{x}^{5}{t}^{2}-36720\,{t}^{2}{{\it d_{1}}}^{2}+66960\,{t}
^{2}{{\it c_{1}}}^{2}-8640\,{x}^{2}{{\it c_{1}}}^{2}-350280\,{t}^{4}{x}^{3}+
436068\,{t}^{5}{x}^{2}+124800\,{t}^{7}x-60672\,{t}^{5}{x}^{3}+140448\,
{t}^{6}{x}^{2}+1313928\,{t}^{5}x-1694730\,{t}^{6}x-83520\,{x}^{3}{t}^{
3}+10368\,{x}^{5}t+36000\,{x}^{4}{t}^{2}-48960\,{t}^{4}{x}^{4}+35280\,
{x}^{4}t+437760\,{t}^{3}x+17280\,t{x}^{3}-218160\,{t}^{2}{x}^{2}-
2115270\,{t}^{4}x-126000\,{t}^{2}{x}^{3}+628200\,{t}^{3}{x}^{2}+2880\,
{\it d_{1}}\,{x}^{4}-8640\,{{\it d1}}^{2}t-8640\,{x}^{2}{{\it d_{1}}}^{2}-
532440\,{x}^{2}{t}^{4}-234720\,{t}^{4}{\it c_{1}}-111852\,{t}^{5}{\it d_{1}}
+129564\,{t}^{5}{\it c_{1}}+8064\,{x}^{5}{\it c_{1}}-432000\,{\it c_{1}}\,{t}^{
3}-73440\,t{{\it c_{1}}}^{2}+8640\,x{{\it c_{1}}}^{2}-29160\,{t}^{2}{\it d_{1}}
+244080\,{t}^{2}{\it c_{1}}+8640\,{x}^{2}{\it c_{1}}+16740\,t{\it d_{1}}-15660
\,t{\it c_{1}}+3240\,x{\it c_{1}}-29160\,{t}^{3}{\it d_{1}}+985140\,{t}^{4}{
\it d_{1}}+276720\,{x}^{4}{t}^{3}+4320\,{x}^{2}{\it d_{1}}-145152\,{x}^{6}{t
}^{3}+277440\,{x}^{5}{t}^{4}-6912\,{x}^{8}t+41472\,{x}^{7}{t}^{2}+
168480\,{t}^{3}{{\it d_{1}}}^{2}-121248\,{t}^{5}{x}^{4}-878176\,{t}^{6}{x
}^{3}+11520\,{x}^{3}{{\it d_{1}}}^{2}+2523600\,{t}^{7}{x}^{2}-3288750\,{t
}^{8}x-123552\,{t}^{6}{\it d_{1}}-1078536\,{t}^{6}{\it c_{1}}-4608\,{x}^{6}{
\it c_{1}}-1080\,x{\it d_{1}}-8640\,{x}^{3}{\it d_{2}}-2160\,x{\it d_{2}}+1080\,{
\it c_{1}}-1080\,{\it d_{2}}+1080\,{x}^{2}+53190\,{t}^{2}+2520\,{x}^{3}-
170325\,{t}^{3}-84480\,{t}^{3}{x}^{3}{\it d_{1}}-718560\,{t}^{4}{x}^{2}{
\it c_{1}}+311040\,{t}^{3}{x}^{3}{\it c_{1}}+448704\,{t}^{5}x{\it d1}+
1315872\,{t}^{5}x{\it c_{1}}-9216\,{x}^{5}t{\it d_{1}}-241920\,{t}^{4}{x}^{2
}{\it d_{1}}+69120\,{x}^{4}{t}^{2}{\it d_{1}}+41472\,{x}^{5}t{\it c_{1}}-155520
\,{x}^{4}{t}^{2}{\it c_{1}}-51840\,t{{\it d_{1}}}^{2}{x}^{2}-60480\,{t}^{2}{
{\it d_{1}}}^{2}x+5760\,{x}^{3}t{\it d_{1}}-25920\,x{\it c_{1}}\,t-21600\,x{
\it d_{1}}\,t-108000\,{x}^{2}t{\it d_{1}}+69120\,{x}^{2}t{\it c_{1}}-262440\,{t
}^{4}x{\it c_{1}}-8640\,{x}^{4}t{\it d_{1}}-48960\,{x}^{4}t{\it c_{1}}+538560\,
{\it c_{1}}\,{t}^{3}x+11520\,{\it c_{1}}\,t{x}^{3}-190080\,{\it c_{1}}\,{t}^{2}
{x}^{2}-8640\,t{{\it c1}}^{2}x+60480\,x{{\it d1}}^{2}t-943200\,x{t}^{3
}{\it d_{1}}+438480\,x{t}^{2}{\it d_{1}}+73440\,{x}^{2}{t}^{2}{\it d_{1}}+
358920\,{t}^{4}x{\it d_{1}}+182880\,{t}^{3}{x}^{2}{\it c_{1}}-31680\,{t}^{2}
{x}^{3}{\it d_{1}}+66240\,{t}^{2}{x}^{3}{\it c_{1}}-27360\,{t}^{3}{x}^{2}{
\it d_{1}}+644130\,{t}^{4}+1440\,{x}^{4}+1790073\,{t}^{5}-4320\,{x}^{5}-
2688\,{x}^{6}+26250\,{t}^{8}-8682\,{t}^{6}+2125893\,{t}^{7}-1536\,{x}^
{8}-384\,{x}^{7}+2160\,{{\it d_{1}}}^{2}+2160\,{{\it c_{1}}}^{2}+1828125\,{t
}^{9}+512\,{x}^{9}-35640\,{t}^{2}{\it d_{2}}-4320\,{x}^{2}{\it c_{2}}-2880\,
{x}^{3}{\it c_{2}}+8640\,{x}^{3}{\it d_{1}}+5760\,{x}^{5}{\it d_{1}}+16200\,t{
\it d_{2}}-7560\,t{\it c_{2}}+2160\,x{\it c_{2}}+5760\,{x}^{4}{\it c_{2}}-87480\,{
t}^{2}{\it c_{2}}-216000\,{t}^{4}{\it d_{2}}+65880\,{t}^{3}{\it c_{2}}-63000\,{
t}^{4}{\it c_{2}}+65160\,{t}^{3}{\it d_{2}}-69120\,{x}^{2}t{\it c_{1}}\,{\it d_{1}
}+207360\,{t}^{2}{\it c_{1}}\,x{\it d_{1}}-60480\,ixt{{\it d_{1}}}^{2}-51840\,x
t{\it c_{1}}\,{\it d_{1}}-63360\,{t}^{3}{\it d_{1}}\,{\it c_{1}}+17280\,{x}^{2}{
\it c_{1}}\,{\it d_{1}}+8640\,x{\it c_{1}}\,{\it d_{1}}+12960\,t{\it c_{1}}\,{\it c_{2}}
+247680\,{t}^{3}x{\it d_{2}}-30240\,{t}^{2}{\it c_{1}}\,{\it d_{1}}+8640\,i{x}^
{2}{{\it d_{1}}}^{2}-8640\,i{x}^{2}{{\it c_{1}}}^{2}+23040\,{x}^{3}t{\it d_{2}}
-36720\,x{t}^{2}{\it c_{2}}-4320\,xt{\it d_{2}}+21600\,{x}^{2}t{\it d_{2}}-
34560\,{x}^{3}t{\it c_{2}}+30240\,xt{\it c_{2}}+30240\,{x}^{2}t{\it c_{2}}-
103680\,{t}^{2}{x}^{2}{\it d_{2}}+77760\,{t}^{2}{x}^{2}{\it c_{2}}-77760\,{t
}^{3}x{\it c_{2}}+12960\,t{\it d_{1}}\,{\it d_{2}}+38880\,{\it c_{1}}\,t{\it d_{1}}-
41040\,{t}^{2}x{\it d_{2}}+760320\,i{x}^{3}{t}^{3}{\it d_{1}}-21600\,i{x}^{2
}t{\it c_{2}}-14400\,it{x}^{4}{\it d_{1}}-64800\,ix{t}^{2}{\it d_{1}}-36720\,ix
{t}^{2}{\it d_{2}}+56160\,ixt{\it c_{1}}+77760\,i{t}^{2}{x}^{2}{\it d_{2}}+8640
\,ix{\it d_{1}}\,{\it c_{2}}+4320\,ixt{\it c_{2}}-1440\,i{x}^{4}+1080\,i{x}^{2}
+33750\,i{t}^{2}+1152\,i{x}^{6}+2128626\,i{t}^{6}+327870\,i{t}^{4}+
1261875\,i{t}^{8}+768\,i{x}^{8}-69120\,it{\it c_{1}}\,x{\it d_{1}}+4320\,i{{
\it d_{1}}}^{2}x+4320\,i{x}^{2}{\it c_{2}}+8640\,i{\it d_{1}}\,{{\it c_{1}}}^{2}+
228960\,i{t}^{4}{\it d_{1}}+23760\,it{{\it c_{1}}}^{2}+62640\,i{{\it d_{1}}}^{2
}t+35640\,i{t}^{2}{\it c_{2}}+5760\,i{x}^{4}{\it d_{2}}+216000\,i{t}^{4}{
\it c_{2}}+63360\,i{t}^{3}{{\it c_{1}}}^{2}+286200\,i{t}^{2}{\it d_{1}}+4320\,i
{x}^{2}{\it d_{1}}+226304\,i{x}^{6}{t}^{3}+1582920\,i{t}^{4}{x}^{3}+32940
\,i{x}^{2}t+315000\,i{t}^{3}{x}^{2}+2292864\,i{t}^{5}{x}^{4}+112608\,i
{x}^{5}{t}^{2}+3072\,i{x}^{8}t+4795200\,i{t}^{7}{x}^{2}+1102890\,i{t}^
{6}x+19440\,i{x}^{4}t-23040\,i{x}^{3}t{\it c_{2}}-315360\,i{\it c_{1}}\,{t}^
{2}{x}^{2}-195840\,i{x}^{4}{t}^{2}{\it d_{1}}-168480\,i{t}^{3}{\it d_{1}}\,{
\it c_{1}}-207360\,i{t}^{2}{{\it c_{1}}}^{2}x-8640\,i{\it c_{1}}\,x{\it d_{2}}-
241920\,i{x}^{2}{t}^{4}{\it c_{1}}-9216\,i{x}^{5}{\it c_{1}}\,t-34560\,it{x}
^{3}{\it d_{2}}-247680\,i{t}^{3}x{\it c_{2}}-11520\,i{\it c_{1}}\,{x}^{3}{\it
d_{1}}-64800\,it{\it c_{1}}\,{\it d_{1}}-1477440\,i{x}^{2}{t}^{4}{\it d_{1}}-84480
\,i{x}^{3}{t}^{3}{\it c_{1}}-17280\,it{\it d_{1}}\,{\it d_{2}}-17280\,it{\it c_{1}
}\,{\it c_{2}}-155520\,i{x}^{2}{t}^{2}{\it d_{1}}-12960\,ix{\it d_{1}}\,t-12960
\,it{\it d_{1}}\,{\it c_{2}}-77760\,i{t}^{3}x{\it d_{2}}-4320\,ix{{\it c_{1}}}^{2}
-128250\,ix{t}^{2}-1392210\,i{t}^{4}x-36864\,i{x}^{7}{t}^{2}-875520\,i
{x}^{5}{t}^{4}-4108032\,i{t}^{6}{x}^{3}-585840\,i{x}^{4}{t}^{3}-14016
\,i{x}^{6}t-1901124\,i{t}^{5}{x}^{2}-43920\,i{t}^{2}{x}^{3}-3240000\,i
{t}^{8}x-3072\,i{x}^{6}{\it d_{1}}-4320\,i{x}^{2}{\it c_{1}}-4320\,i{x}^{2}{
\it d_{2}}-63000\,i{t}^{4}{\it d_{2}}-160920\,i{t}^{2}{\it c_{1}}-2880\,i{x}^{4
}{\it c_{1}}-774900\,i{\it c_{1}}\,{t}^{4}-87480\,i{t}^{2}{\it d_{2}}-796464\,i
{t}^{6}{\it d_{1}}-123552\,i{t}^{6}{\it c_{1}}+17280\,i{\it c_{1}}\,{x}^{3}t+
12960\,i{\it c_{1}}\,t{\it d_{2}}+448704\,i{t}^{5}x{\it c_{1}}+69120\,i{x}^{2}{
{\it c_{1}}}^{2}t+1080\,i{\it c_{2}}+8640\,i{{\it d_{1}}}^{3}+69120\,i{x}^{4}{t
}^{2}{\it c_{1}}+384\,i{x}^{7}+540\,i{\it c_{1}}+1080\,i{\it d_{1}}+30240\,ixt{
\it d_{2}}+51840\,i{x}^{2}{\it d_{1}}\,t{\it c_{1}}+60480\,i{t}^{2}{\it d_{1}}\,x{
\it c_{1}}+8640\,ix{\it c_{1}}\,{\it d_{1}}-4320\,{x}^{2}{\it d_{2}}+4320\,{\it c_{1}
}\,{\it c_{2}}+4320\,{\it d_{1}}\,{\it d_{2}}-3960\,i{x}^{3}-270\,ix-3744\,i{x}
^{5}-1094349\,i{t}^{7}+8640\,{\it c_{1}}\,{{\it d_{1}}}^{2}+2160\,ix{\it c_{2}}
+2880\,ix{t}^{3}{\it d_{1}}+687500\,i{t}^{9}+457065\,i{t}^{3}+1135251\,i{
t}^{5}+1485\,it+135\,i-8640\,x{\it d_{1}}\,{\it d_{2}}-8640\,x{\it c_{1}}\,{
\it c_{2}}-17280\,t{\it d_{1}}\,{\it c_{2}}+17280\,{\it c_{1}}\,t{\it d_{2}}+1236960
\,i{\it c_{1}}\,{t}^{3}x+103680\,i{t}^{2}{x}^{2}{\it c_{2}}-4320\,i{\it d_{1}}
\,{\it c_{2}}-6480\,i{t}^{2}{x}^{2}-563400\,i{x}^{2}{t}^{4}-1802304\,i{t}
^{5}{x}^{3}-832104\,i{t}^{5}x-15360\,i{x}^{7}t-89280\,i{t}^{3}x-
2036400\,i{t}^{7}x-1080\,ixt-398592\,i{x}^{5}{t}^{3}-5760\,i{x}^{5}t-
18720\,i{x}^{4}{t}^{2}-336852\,i{t}^{5}{\it c_{1}}-8100\,it{\it c_{1}}-65160
\,i{t}^{3}{\it c_{2}}-2880\,i{x}^{3}{\it d_{2}}-7560\,it{\it d_{2}}-16200\,it{
\it c_{2}}-5760\,i{x}^{5}{\it c_{1}}-8640\,i{\it c_{1}}\,{x}^{3}-149400\,i{\it
c_{1}}\,{t}^{3}-79920\,i{t}^{3}{\it d_{1}}+8640\,{{\it c_{1}}}^{3}+1384128\,ix{
t}^{5}{\it d_{1}}+34560\,it{\it d_{1}}\,{x}^{3} )/(9-612\,xt-192\,{x}
^{3}{\it c_{1}}-22500\,{t}^{5}x-8928\,{x}^{3}{t}^{3}-576\,{x}^{5}t+2928\,
{x}^{4}{t}^{2}+792\,{t}^{3}x+96\,t{x}^{3}-2232\,{t}^{2}{x}^{2}+18300\,
{x}^{2}{t}^{4}-2808\,{\it c_{1}}\,{t}^{3}+864\,t{\it d_{1}}-792\,t{\it c_{1}}+
144\,x{\it c_{1}}-1056\,{t}^{3}{\it d_{1}}+108\,{x}^{2}+2835\,{t}^{2}-1152\,
{x}^{2}t{\it d_{1}}+864\,{x}^{2}t{\it c_{1}}+1008\,{\it c_{1}}\,x{t}^{2}+3456\,
x{t}^{2}{\it d_{1}}+10179\,{t}^{4}+48\,{x}^{4}+64\,{x}^{6}+15625\,{t}^{6}
+144\,{{\it d_{1}}}^{2}+144\,{{\it c_{1}}}^{2})
$.

\end{document}